\pdfoutput=1
\documentclass[preprint,superscriptaddress,amsmath,amssymb,floatfix]{revtex4-1}
\usepackage{manfnt}
\usepackage{graphicx}
\usepackage{dcolumn}
\usepackage{bm}
\usepackage{hyperref}
\usepackage{dcolumn}
\usepackage{tikz}
\usepackage[normalem]{ulem}
\usepackage[makeroom]{cancel}
\usepackage{bbm}
\usepackage[normalem]{ulem}
\hypersetup{unicode=true,
            pdftitle={},
            pdfauthor={},
            pdfborder={0 0 0},
            breaklinks=true}
\tikzset{every label/.style={font=\footnotesize,inner sep=1pt}}

\newcommand{\Rho}{\mathrm{P}}

\begin{document}

\title{Bayesian inference of  polymerase  dynamics over the exclusion process.}
\author{Massimo Cavallaro}
\email{m.cavallaro@warwick.ac.uk}
\affiliation{Mathematics Institute, University of Warwick, Coventry, UK}
\affiliation{School of Life Sciences, University of Warwick, Coventry, UK}
\affiliation{Zeeman Institute for Systems Biology and Infectious Disease Epidemiology Research, University of Warwick, Coventry, UK}
\author{Yuexuan Wang}
\affiliation{Institute of Applied Statistics, Johannes Kepler Universit\"at, Linz, Austria}
\author{Daniel Hebenstreit}
\affiliation{School of Life Sciences, University of Warwick, Coventry, UK}
\author{Ritabrata Dutta}
\affiliation{Department of Statistics, University of Warwick, Coventry, UK}

%\date{\today}

\begin{abstract}
Transcription is a complex phenomenon that permits the conversion of genetic information into phenotype
by means of an enzyme called  RNA polymerase, which erratically moves along and scans the DNA template.
We perform Bayesian inference over a paradigmatic mechanistic model of non-equilibrium statistical physics,
i.e., the asymmetric exclusion processes in the hydrodynamic limit, assuming a Gaussian process
prior for the  polymerase progression rate as a latent variable.
Our framework allows us to infer the speed of polymerases during transcription given their spatial distribution,
whilst avoiding the explicit inversion of the system's dynamics.
The results, which show processing rates strongly varying with genomic position and minor role of traffic-like congestion, may have strong implications for the understanding of gene expression.
\end{abstract}

\maketitle

\section{Introduction}
DNA is a long polymeric molecule that encodes information as a sequence of nucleotides (Nts).
Turning this information into a phenotype is a complex phenomenon hinged upon transcription, the molecular process
in which particular segments of DNA (i.e., the genes) are scanned and their information is copied into mRNA by the enzyme RNA polymerase II (PolII). The transcription itself consists of several steps which can be differentially regulated to alter the timing and the output of the mRNA production~\cite{Munsky2012, Rajala2010}.

The transcription can also be seen as a non-equilibrium process, where the PolIIs are being transported as particles on a one dimensional lattice, the lattice being the DNA template which the PolIIs bind to. We can further consider this process having  left and right boundaries, representing the transcription start site (TSS) and the transcription end site (TES), respectively (Fig.~\ref{fig:1}~A).
Within the gene body, the PolIIs erratically travel along the template and their abrupt slowing down in certain genomic regions is known as \textit{pausing} dynamics~\cite{Jonkers2015, Mayer2017}.
While the pausing is an essential part of the transcriptional machinery and contributes to the regulation of genes’ expression levels, a comprehensive  quantitative understanding of its dynamics is still missing~\cite{Adelman2012, Liu2015}.

We present a modelling framework to help understand gene regulation and quantitatively study the pausing dynamics given real-world data. In literature,
a number of different mechanistic models have been introduced to elucidate
transcription, starting from the simple telegraph model~\cite{peccoud1995markovian} to more complicated multi-state 
models that account for many interactions~\cite{Tripathi2008, Dobrzynski2009, Cao2020, Szavits-Nossan2022}, 
with each model reflecting determinate aspects of the whole biological system complexity.
Here we are primarily interested in the pausing and employ a generalisation of a paradigmatic model of particle transport, the asymmetric simple exclusion process (ASEP,~\cite{MacDonald1968,MacDonald69,Spitzer1970}) in the hydrodynamic limit~\cite{Benassi1987}.
The ASEP is a class of models of particles on a one-dimensional lattice, whose  behaviour is chiefly determined by the rates at which the particles hop on the lattice.
More specifically we require the rate profile function, which we refer to as $\tilde{p}$,
to be spatially varying yet smooth as in reference~\cite{Stinchcombe2011, Lakatos2006},
see also~\cite{Harris2004},
thus making it possible to model this function by a Gaussian process (GP)~\cite{Rasmussen2006}.
Noticing the analogy between the PolII transport in
the gene body and the particle hopping in the exclusion process, learning $\tilde{p}$ allows the study of the pausing dynamics in a gene.
Importantly, we provide an inferential scheme to learn this rate function by Bayesian inference given real molecular biology data, assuming  a  prior  on  the  profile  function  induced  by a GP prior on a latent variable.
In other words, integrating the dynamics defined by the rate $\tilde{p}$ generates transient time-course density profiles;
we estimating $\tilde{p}$ given observed density profiles without explicitly inverting
the system's dynamics.
Other models of PolII dynamics also leverage GPs for inference from biological data, with GPs representing transcriptional activity over time~\cite{WaMaina2014, Honkela2015}. In contrast, the GP here describes a function of genomic position, with its minima corresponding to pausing regions.
Due to its generality, our framework can be  deployed to estimate the rate profiles of any one-dimensional transport problem.

The manuscript is organised as follows.
Section \ref{sec:bio} describes the biology of pausing
and the next-generation sequencing (NGS) data types which are available to study it.
Section \ref{sec:model} and \ref{sec:stat}, respectively, discuss the asymmetric simple exclusion process as a
mathematical model for transcription with pausing and a Bayesian inferential framework for model fitting.
We present the results in section \ref{sec:results} and conclude with a discussion in section~\ref{sec:discussion}.

\section{Model definition}
\subsection{Biological processes and data}
\label{sec:bio}
RNA polymerases have a central role in the biology of transcription.
We distinguish different classes of RNA polymerases, each
having  different structure  and control mechanism.
Bacteria and Archaea only have one RNA polymerase type. Eukaryotes have multiple types, of which RNA polymerase II (PolII) is known to catalyse synthesis of protein-encoding RNA (messenger RNA or mRNA).
In this paper we describe mRNA transcription by PolII, 
but the inferential framework we present is general and can 
be extended to other transport phenomena.
PolII binds to DNA upstream the TSS,
initiates the mRNA synthesis, and then traverses the
DNA downstream (elongation) until it pauses at a certain gene location,
ready to respond to a developmental or environmental signal that instructs to
resume the elongation.
PolIIs are also found proximal to TSS in a so-called ``poised'' state, which has not initiated synthesis of the mRNA chain. Poised and paused PolII can be differentiated as only paused PoIIs have a tail of nascent mRNA and is bound to transcription factor Spt5~\cite{Erickson2018}.
The process terminates when the PolII reaches the TES and the transcribed mRNA is released.
As a result of these steps, the output is modulated in both timing and intensity.
However, many details, such as the pausing, are not well understood~\cite{Adelman2012}.
The presence of transcriptional pausing   in Eukaryotes is revealed by several assays based on NGS,
which is widely used in molecular biology to study molecules involved in genic processes.
In the PolII \textit{ChIP-seq} assay, PolII-bound DNA is isolated by chromatin immunoprecipitation with a PolII antibody and is
then subject to high-throughput sequencing.
This provides a genome-wide view of the
PolII binding sites for all forms of PolII, including both those poised or transcriptionally engaged and those which are bound to DNA and static. In ChIP-seq experiments, DNA fragments extracted from cells and associated with a specific protein (here polymerase) are amplified, sequenced, and mapped to the reference genome, with fragments generally in the 150-300 Nt range \cite{Park2009} (while transcribing PolII covers less than 50 Nts of DNA~\cite{Ehara2017}). This means that the precise locations of the individual proteins are not known and the assay only returns the overlap of reads from many different cells.
For each genomic position, PolII ChIP-seq returns a signal as a proxy of polymerase occupancy.

\begin{figure*}
    \includegraphics[width=0.95\linewidth]{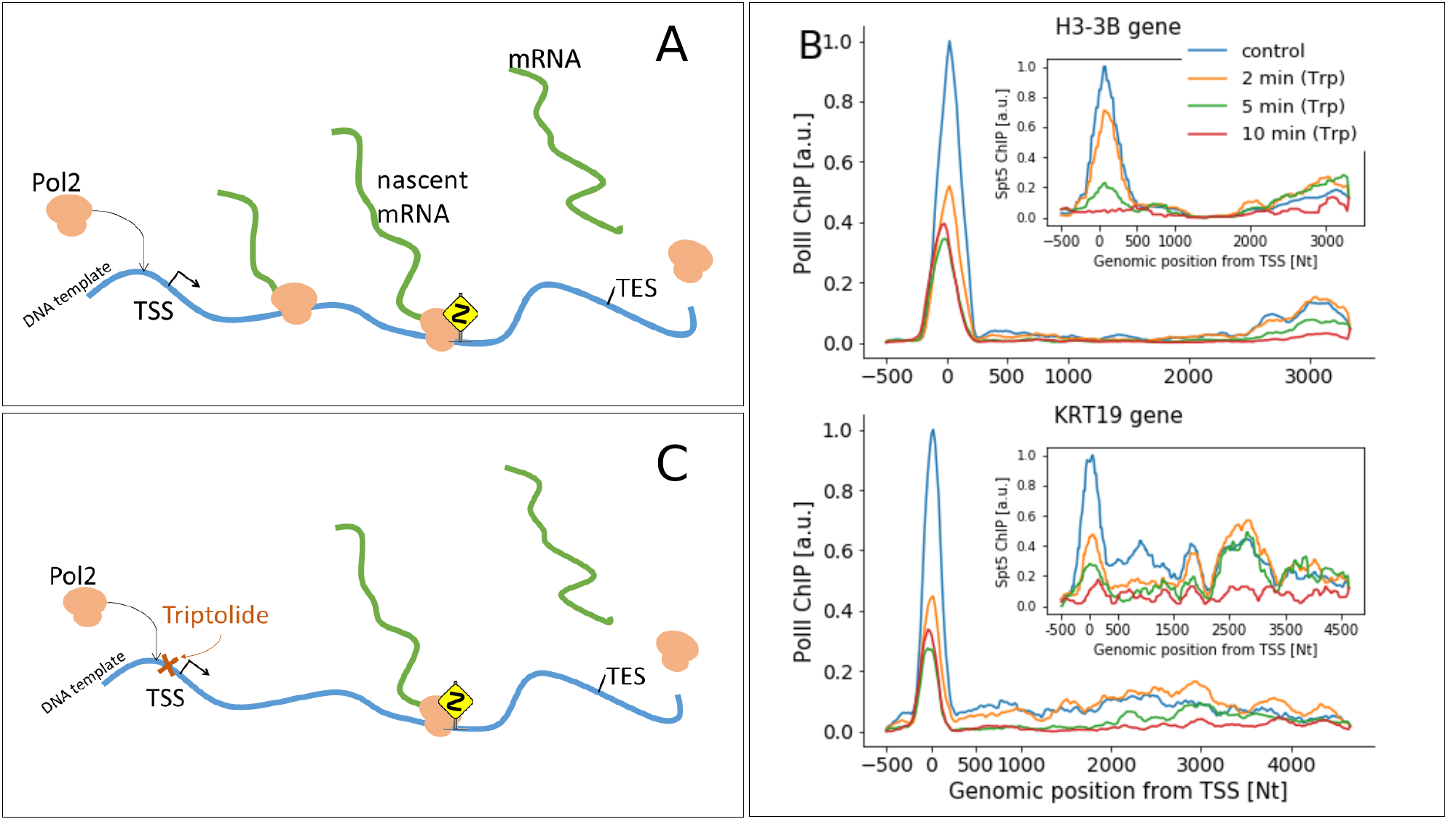}
    \caption{\label{fig:1} Biological processes and data.
    A) Simplified diagram of mRNA synthesis. PolII molecules bind to the DNA upstream of transcription start site (TSS) and moves downstream towards the transcription end site (TES), where it is released along with the  synthesized mRNA.
    In certain genomic regions (indicated by a dangerous bend sign), PolIIs slow down.
    B) ChIP-seq experiments yield the relative abundance of PolII at each genomic position, here illustrated for H3-3B  (top) and
    KR19  (bottom) genes; insets show the Spt5-bound PolII abundances for the same genes. C) In the presence of triptolide, transport is blocked upstream of TSS, while transcriptional engaged PolII are allowed to complete elongation; this is reflected in the ChIP-seq profiles obtained 2, 5, and 10 minutes after treatment (also in B).
    }
\end{figure*}

For this study, we binned ChIP-seq reads from genomic ranges of selected genes (from cultured human cell lines, \textit{Materials and Methods}) into 20-Nt bins,
thus yielding coarse-grained read profiles (which we refer to as $y$) 
such as those
illustrated in Fig.~\ref{fig:1}~B.
The number of these reads at a position $x$ is proportional to the occupation probability $\varrho(x)$.
The proportionality factor, which depends on the number of cells used in the experiment and on further signal amplification intrinsic to the sequencing procedure, cannot be directly accessed with precision and is only known with substantial uncertainty~\cite{Hu2015}.

Other methods available to study the pausing include but are not limited to \textit{NET-seq},
where nascent mRNA chunks associated with immunoprecipitated PolII complexes are
isolated and sequenced~\cite{Nojima2015}, \textit{GRO-seq}, where RNAs recently transcribed only by
transcriptionally-engaged PolIIs are sequenced~\cite{Core2008}, and \textit{PRO-seq}, which is similar
to GRO-seq but reaches single-nucleotide  resolution~\cite{Kwak2013}.
The evidences of PolII transport are particularly clear in time-course experiments, where sequencing
data are collected over time following a perturbation. As an example, time-variant PRO-seq has been suggested to 
estimate pausing times in key peak regions~\cite{Zhang2021}.
A classical way to perturb these molecular dynamics is inhibiting the initiation by treating the cells
with triptolide (Trp), which is a highly specific drug that blocks initiation~\cite{Jonkers2014, Erickson2018}.
This permits the PolII already engaged in transcription to progress further downstream the gene
while new PolIIs are prevented to attach, thus freeing upstream genomic regions as the run-on time
progresses (Fig.~\ref{fig:1} C). Our approach consists of using the read profiles $y$ as functions of $x$, collected at fixed run-on
times $t_1$, $t_2$, $t_3$, and $t_4$ after treatment, to infer the dynamics. While Trp inhibits new initiation, poised PolII upstream the TSS can still pass through it, enter
the gene template, and perform elongation immediately after Trp treatment~\cite{Jonkers2014, Erickson2018}.
To account for this, we also perform inference over Spt5 ChIP-seq data, where the poised polymerases are masked
while those bound are detected~\cite{Erickson2018}.

These types of experiments reveal the presence of a flux of PolIIs, which is the signature of the
non-equilibrium physics involved in the elongation process.
The profile $y^*$ observed prior to the treatment corresponds to a non-equilibrium stationary state (NESS).
Disrupting initiation with Trp yields a transient state, which evolves from $y^*$ until it settles down to a new NESS.

\subsection{Mathematical model}
\label{sec:model}
The transport of particles on a one dimensional lattice is a  well-studied problem in
mathematics and physics. Its basic features are captured by the asymmetric simple exclusion
process (ASEP)~\cite{Spitzer1970}, which defines the stochastic dynamics of interacting particles on a discrete lattice,
which we take here to be a one-dimensional chain with open boundary conditions.
Let the total number of lattice sites be $N$.
The state of each site $i$, $1 \le i\le N$, is characterized by the occupation number
$n_i$ such that $n_i = 0$ if the site is empty and $n_i = 1$ if it is occupied by a particle.
The evolution proceeds in continuous time.
A particle on site $i<N$ hops rightward into the site $i+1$ with rate $p_i$,
the transition being successful only if the site $i+1$ is empty.
Similarly, a particle on site $i > 1$ hops leftward into $i-1$ with rate $q_{i}$, if the site $i-1$ is empty.
Further, particles on the left (right) boundary site  $i=1$ ($i=N$) leave the lattice at rate $q_1$ ($p_N$),
while particles are injected in the same boundary site at rate $p_0$ ($q_{N+1}$) if the site is empty.
The constraint that a jump can occur only if the target is empty prevents the accumulation of more
than one particle on a site and is generically referred to as the exclusion rule.
This rule allows particle collision, which
causes congestion when the particle density is sufficiently high and
permits phase transitions between a low density, high density, and a
maximum current phase even if the systems is one dimensional~\cite{Chou2011}.
Interestingly, based on theoretical considerations, it has been suggested that traffic-like congestion of PolIIs is important in transcription~\cite{Klumpp2008, Cholewa-Waclaw2019, Tripathi2009}.

While the ASEP was originally proposed to model biopolymerization on  on nucleic acid templates~\cite{MacDonald1968, MacDonald69},
this and related models have  been more recently applied to diverse problems,  including protein translation~\cite{Zia2011, Szavits-Nossan2018,Erdmann-Pham2019},
but also, e.g.,  
molecular motors~\cite{Lipowsky2001}  and pedestrian and vehicle traffic~\cite{Chowdhury2000}.
Applications to transcription incorporating disordered dynamics and obstacles (e.g., \cite{Wang2014, Waclaw2019}) were also proposed.
ASEP's theoretical appeal is due to its analytical results representative of a large class of models~\cite{Kardar1986, Bertini1997}
and a convenient mean-field treatment that yields the exact stationary solution~\cite{Derrida1993}.
In the context of transcription, particles entering site $1$,
moving along the chain, and exiting from site $N$ correspond to initiation,
elongation, and termination, respectively.
In our setting, the lowest values of $p_i$ 
correspond to genomic locations were elongation slows down.

The dynamics of the expected occupation of a single site $i$ in the bulk are governed by
the lattice continuity equation
\begin{equation}
    \frac{d}{dt}  \mathbb{E} (n_i(t)) = J^{\mathrm{left}}(t) - J^{\mathrm{right}}(t),  \label{eq:expected_n} 
\end{equation}
$0<i<N$,    where $\mathbb{E}$ denotes expectation value and $J^{\mathrm{left}}(t)$ and $J^{\mathrm{right}}(t)$ are the average flux of particles from site $i-1$ to site $i$ and from site $i$ to site $i + 1$, respectively.
These are subject to the exclusion rule and therefore obey
\begin{equation}
\begin{aligned}
    J^{\mathrm{left}}(t)  &= p_{i-1} \mathbb{E} (n_{i-1}(t) (1-n_{i}(t))) \\
                          &- q_i \mathbb{E} ( n_i (t)(1- n _{i-1}(t)) ),\\
    J^{\mathrm{right}}(t) &= p_{i}  \mathbb{E} (n_{i}(t) (1-n_{i+1}(t)) )  \\
                          &- q_{i+1} \mathbb{E} ( n_{i+1} (t)(1- n _{i}(t)) ) .
    \label{eq:J}
\end{aligned}
\end{equation}
In order to exactly solve these dynamics, 
second-order moments
such as $\mathbb{E} (n_i(t) n_{i+1}(t) )$ 
need 
to be known. Under independence assumption, these moments are factorised, which in our case amounts to replacing equations~\eqref{eq:expected_n}--\eqref{eq:J} with
\begin{equation}
\begin{aligned}
    \frac{d}{dt}  \phi_i(t) &= %J_{i-1,i}(t)-J_{i,i+1}(t)
    p_{i-1} \phi_{i-1}(t)  (1 - \phi_{i}(t)) - p_i\phi_i(t) (1 - \phi_{i+1}(t))\\
    &+q_{i+1} \phi_{i+1}(t) (1 - \phi_{i}(t))  - q_i \phi_i (t) (1 - \phi_{i-1}(t)),
       \label{eq:mean_field}
\end{aligned}
\end{equation}
where we used $\phi_i(t) := \mathbb{E} (n_i(t))$ to lighten the notation.
In other words, equations \eqref{eq:mean_field} define the so-called mean-field dynamics of the asymmetric exclusion process, which are known to approximate well the true dynamics in many contexts, predict crucial features such as dynamical phase-transitions, and ease mathematical treatment~\cite{Derrida1993, Chou2011, Lazarescu2015}.
With open boundaries,
\begin{align}
\begin{aligned}
    \frac{d}{dt} \phi_1(t) &= p_0 \, (1 - \phi_1(t)) - p_1 \phi_1(t) (1 - \phi_2(t)) \\
    & - q_1 \phi_1(t)  + q_2 \phi_2(t) (1 - \phi_1(t)),
\end{aligned} \label{eq:openboundary1} \\ 
\begin{aligned}
    \frac{d}{dt} \phi_N(t) &= p_{N-1} \phi_{N-1}(t) (1 - \phi_N(t)) - p_N \phi_N(t) \\
    & + q_{N+1} \, (1 - \phi_{N}(t) ) - q_{N} \phi_{N} (1 - \phi_{N-1}).
\end{aligned}    \label{eq:openboundary2}
\end{align}

To match the available data that is coarse grained (Fig. \ref{fig:1}~B),
instead of considering particles individually we rely on their hydrodynamics description,
which is obtained as follows. We assume Euler scaling with constant $a$
and let $a \to 0, N \to \infty$, with $L:= N a$ held finite.
We define the functions $\varrho:\mathbbm{R}^2 \to \mathbbm{R}^+_0$, $\tilde p:\mathbbm{R}\to\mathbbm{R}^+_0$, and $\tilde q:\mathbbm{R}\to\mathbbm{R}^+_0$
such that
they are analytic and bounded on $]0,L[ \times ]0,\infty[$, $]0,L[$, and $]0,L[$, respectively,
and
\begin{equation}
\begin{aligned}
\phi_{i}(t) & = \varrho((i-1) \, a,t),\\
 a \, p_i & = \tilde p((i-1) \, a),\\
 a \, q_i & = \tilde q((i-1) \, a).
\end{aligned}
\end{equation}
We further assume that the left and right jump rates satisfy
$\tilde{q}(x) = b \,\tilde{p}(x)$, $\forall x \in [0,L]$,
with $0 \le b < 1$,  where $b$ governs the relative strength of the
non-equilibrium driving forces.
The case $b=0$ corresponds to a \emph{totally asymmetric exclusion process} (TASEP), while the limit case
$b=1$ corresponds to the \emph{symmetric exclusion process}. Intermediate values $0<b<1$ correspond to
settings where the particles can jump in both directions, but are
driven rightwards on average.
A continuum-limit counterpart of equations~\eqref{eq:J}, as derived in references~\cite{Harris2004, Stinchcombe2011},
is
\begin{equation}
J(x,t) =  \tilde p(x) \varrho(x - a/2,t)(1-\varrho(x+a/2,t)) 
       -   \tilde q(x) \varrho(x + a/2,t)(1-\varrho(x-a/2,t)),
\end{equation}
which, using first-order Taylor expansion, yields
\begin{equation}
J(x,t) \approx  (\tilde p(x) - \tilde q(x))\varrho(x,t) (1 - \varrho(x,t)) 
-  \frac{a}{2}  (\tilde p(x) + \tilde q(x)) \frac{\partial}{\partial x} \varrho(x,t).
      \label{eq:J_expansion}
\end{equation}
To lighten the mathematical notation, we define the two quantities
\begin{equation}
\begin{aligned}
\lambda(x) &:= ( \tilde p(x) - \tilde q(x)) =  \tilde p(x) (1-b), \\
\nu(x) &:= \frac{a}{2} (\tilde p(x) + \tilde q(x)) = \frac{a}{2} \tilde p(x) (1+b);
      \label{eq:lambdanu}
\end{aligned}
\end{equation}
their ratio is constant in $x$, viz.,
${\nu(x)}/{\lambda(x)} = {a}/{2} \, (1+b)/(1-b)$,
which equals $a/2$ in the totally asymmetric case.

Substituting \eqref{eq:J_expansion}--\eqref{eq:lambdanu} into the continuity equation
\begin{equation}
  \frac{\partial}{\partial t} \varrho(x,t) = - \frac{\partial}{\partial x} J (x,t),
\end{equation}
which is the hydrodynamics limit of equation~\eqref{eq:expected_n},
gives the non-linear partial differential equation
\begin{equation}
  \frac{\partial }{\partial t} \varrho(x,t) = - \frac{\partial}{\partial x} \Big\{ \lambda(x) \varrho(x,t) (1 - \varrho(x,t)) %\\
  - \nu(x) \frac{\partial}{\partial x} \varrho(x,t) \Big\},
  \label{eq:non_linear_PDE}
\end{equation}
which can be linearised to
\begin{equation}
   \frac{\partial }{\partial t} u(x,t) =   \frac{\tilde{p}(x)} {2}
  \Big\{{a^2} (1+b)   \frac{\partial^2}{\partial^2 x}u (x,t) \\-  \frac{(1 - b)^2}{1 + b} u(x,t) \Big\}
  \label{eq:linear_PDE}
\end{equation}
by means of a generalisation of the Cole-Hopf transform (Appendix~\ref{app:Cole-Hopf_transform}
and references~\cite{Hopf1950, Cole1951, Harris2004}).

In transcription, the particle flux is left to right. While PolIIs can backtrack few Nts under certain circumstances~\cite{Nudler1997, Julicher1998, Wang2009}, this phenomenon is overall minor and is not observable at our ChIP-seq resolution. Therefore, we assume $b=0$ and focus on the inference of the net forward rate profile $\tilde{p}(x)$. For simplicity we we also set $a=1$,
arguing that our considerations remain valid with such a choice. 
The required boundary values $\varrho(0,t)$,  $\varrho(L,t)$, and
 $\varrho(x,0)$, and the numerical scheme used to integrate equation~\eqref{eq:linear_PDE} are detailed in Appendices~\ref{app:Cole-Hopf_transform} and~\ref{app:Numerical_integration}.

\begin{figure*}[!t]
    \includegraphics[width=0.99\linewidth]{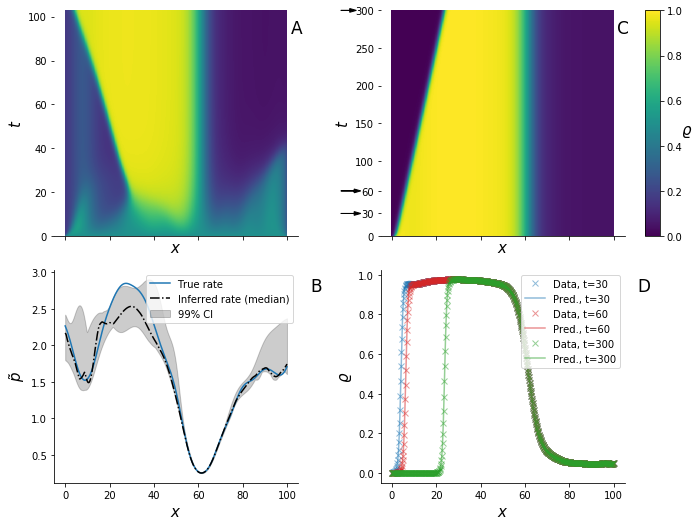}
    \caption{\label{fig:NESS}
    Simulation study. A) A non-equilibrium stationary state  (NESS) of profile $\varrho^*(x)$ is obtained integrating the hydrodynamic TASEP with open boundaries, initial density profile $\varrho(x,0)=0.5$ $\forall x \in [0,L]$, and chosen rate profile (solid line in (B)).
    B) True rate profile (solid line) and inferred rate profile (dash-dot line); the shaded area is $99\%$ credible interval (CI).
    C) Integrating the same dynamics with initial profile $\varrho^*(x)$ and no-influx boundary conditions show that the density decreases in proximity of the left-boundary, similar to ChIP-seq readings followed by Trp treatment; the density profiles corresponding to times $30$, $60$, and $300$ and used for inference are marked by arrows.
    D) The posterior predictive samples (solid lines) are in excellent agreement with the extracted density profiles (cross markers); the posterior predictive dispersion is
    of the order of the line width, see also Fig.~\ref{fig:posterior_zoomin}.  }
\end{figure*}

Integrating equation \eqref{eq:non_linear_PDE} with boundary conditions analogous
 to equations~\eqref{eq:openboundary1}--\eqref{eq:openboundary2} and initial density
 $\varrho(x,0) > 0$ yields a NESS for large $t$,
characterised by a non-vanishing average flux and a density profile $\varrho^*(x)$ which is invariant in time.
Setting the latter as initial condition and further integrating with no inward particle flux ($p_0=q_N=0$)
produces a transient state that mimics the evolution of the PolII profile after Trp treatment until
the density profile vanishes. This is illustrated, for a choice of boundary values and jump rate profile, in Fig.~\ref{fig:NESS}, which also includes the result of the inference process described in the next subsections.

\subsection{Bayesian framework}
\label{sec:stat}
We fit the model to real-world data by means of 
 a Bayesian approach,
leveraging 
its ability to explicitly encode prior hypotheses about the quantities we wish to infer~\cite{gelman2013bayesian}.
We are interested in the forward rate profile $\tilde{p}$.  As this is required to be analytic and non-negative,
it is convenient to assume a Gaussian process (GP) \cite{Rasmussen2006} functional prior on a latent variable $f$ and induce a prior on $\tilde{p}$ using
a sigmoid link function of $f$, such that $\tilde{p} = \tilde{p}_{\mathrm{max}}/(1+\exp(-f))$, which further imposes an
 upper bound $\tilde{p}_{\mathrm{max}}$  to $\tilde{p}$.
The GP prior here defines a distribution over real valued $C^1$ functions in $\mathbb{R}$,
where any finite set of function evaluations $f(x)$ has multivariate normal distribution
with mean $m$ and   covariance kernel $k(x,x';\sigma^2_f,l)=\sigma^2_f \exp(-(x-x')^2/(2\, l^2))$, $x,x' \in \mathbb{R}$.
 In practice, the GP is evaluated at the positions $x_i$, $i=1,\ldots,\mathsf{n}$, where is it equivalent by definition to  $\mathbf{f}\sim\mathcal{N}(m,\mathbf{K})$, a
multivariate normal random variable with mean $m$ and covariance matrix $\mathbf{K}(\sigma_f, l)$ induced by the kernel.

The observations are organised into a collection of values $\mathbf{y}=\{y_{ij}\}_{i=1,2,\ldots,\mathsf{n},j=1,2,\ldots,\mathsf{t}}$,
where the subscripts indicate that an observation is taken at position $x_i$ and time $t_{j}$.
As the values of $x_i$ do not necessarily coincide with the bin centers of  ChIP-seq data,
we used simple linear interpolation to estimate the data at intermediate coordinates.
We assume that the observed values depend on a multiplicative factor and also include
an additive error term $\epsilon\sim\mathcal{N}(0,\sigma_\epsilon)$.
This can be written in terms of the equation $\kappa y_{ij} = \varrho(x_i,t_j) + \epsilon$, where $\kappa$ is the inverse of the amplification factor.
The likelihood $P(\mathbf{y}|\mathbf{f},\sigma_\epsilon,\kappa)$ satisfies
\begin{equation}
    \log P(\mathbf{y}|\mathbf{f},\sigma_\epsilon, \kappa) = -\frac{1}{2 \sigma^2_\epsilon} \sum_{i=1}^{\mathsf{n}} \sum_{j=1}^{\mathsf{t}} (\varrho(x_i,t_j; \mathbf{f})- \kappa y_{ij})
    - \frac{n}{2}\log(2\pi\sigma^2_\epsilon),
    \label{eq:loglikelihood}
\end{equation}
where we made explicit that $\varrho$ depends on $\mathbf{f}$.
For the hierarchical parameters $(m,\sigma_\epsilon,\kappa,\sigma_f,l)=:\theta$
we assume a scaled sigmoid Gaussian prior  probability 
$P(m,\sigma_\epsilon,\kappa,\sigma_f,l)$ such that
\begin{equation}
    \theta  = \theta_{\mathrm{min}}  + \frac{(\theta_{\mathrm{max}} - \theta_{\mathrm{min}})}{1+\exp(-\xi)}, \qquad \xi\sim\mathcal{N}(\mu_\xi, \mathbbm{1} \sigma_\xi),
\end{equation}
where $\theta_{\mathrm{min}}:=(m_{\mathrm{min}},{\sigma_{\epsilon}}_{\mathrm{min}},\kappa_{\mathrm{min}},{\sigma_{f}}_{\mathrm{min}},l_{\mathrm{min}})$, $\theta_{\mathrm{max}}:=(m_{\mathrm{max}},{\sigma_{\epsilon}}_{\mathrm{max}},\kappa_{\mathrm{max}},{\sigma_{f}}_{\mathrm{max}},l_{\mathrm{max}})$, and $(\mu_\xi,\sigma_\xi)$ are referred to as hyperparameters.
Prior distributions are chosen to pull MCMC samples away from inappropriate results that are consistent with the likelihood but would not
be consistent with domain knowledge \cite{gelman2013bayesian}. By using scaled sigmoid Gaussian prior probability bounded
by $\theta_{\mathrm{min}}$  and $\theta_{\mathrm{max}}$, we only search for solutions constrained in an appropriate interval~\cite{Tegner2019}.
By virtue of the Bayes theorem the joint posterior probability for $\theta$ and $\mathbf{f}$ satisfies
\begin{equation}
     P(\mathbf{f},m,\sigma_\epsilon,\kappa,\sigma_f,l|\mathbf{y}) \propto P(\mathbf{y}|\mathbf{f},\sigma_\epsilon,\kappa)
     P(\mathbf{f}|m,\sigma_f,l) P(m)P(\sigma_\epsilon)P(\kappa)P(\sigma_f)P(l),
     \label{eq:posterior}
\end{equation}
which we draw random samples from by Markov chain Monte Carlo (MCMC) sampling, more specifically block Gibbs sampling with elliptical slice sampling at each block \cite{Murray2010, Tegner2019} (Appendix~\ref{app:MCMC}).
Equation \eqref{eq:posterior} expresses the distribution of parameters given the observed data $\mathbf{y}$ and completes the definition of the model.
 It is worth noting that evaluating
 the likelihood also requires computing $\varrho$ by integrating equation~\eqref{eq:non_linear_PDE}
with initial condition $\varrho(x,0) = \kappa \, y^*(x)$,
$\forall x \in [0,L]$.

\section{Results}
\label{sec:results}
We first consider simulated data from a given profile of length $L=100$ obtained from GP draw with parameters $(l, \sigma_f, m, \tilde{p}_{\mathrm{max}})=(7.32,0.67,0.29,3)$.
We integrate the dynamics with NESS initial profile (obtained by fixing the boundary conditions to $\varrho(0,t) = \varrho(L,t) = 0.5$, $\forall t$ ) and no-influx boundary conditions (Fig.~\ref{fig:NESS}~A-B).
The chosen rate profile shows a local minimum close to the left boundary, which yields a minor local perturbation in the density, and a global minimum around $x\approx60$, whose effect propagates along the lattice and 
acts as a major bottleneck, which separates a low density phase downstream from a high-density phase upstream. These minima correspond to regions where particles slow down or pause for an exponentially distributed amount of time.
As the particles leave the system through the right boundary
and are not replenished by the influx through the left boundary,
the region upstream the bottleneck is emptied by a reverse wave front.

For the purpose of testing whether we are able to recover the rate profile from time-course observations, we extract density profiles $\mathbf{y}$ at times $(t_1,t_2,t_3)=(30,60,300)$ and set the hyperparameters
$\theta_{\mathrm{min}}$, $\theta_{\mathrm{max}}$, 
and $(\mu_\xi, \sigma_\xi)$ to $(0,0,0.8,0,0)$, $(2,10,1.2,1,10)$, and $(0,1)$, respectively. 
With these settings and data, we generated $10^4$ MCMC samples targeting the posterior \eqref{eq:posterior},
discarding the first $2 \times 10^3$ as burn-in,
demonstrating that the fitting procedure is able to capture the location of  both the major
and minor minima of the generative model, as well as the overall elongation rate (Fig.~\ref{fig:NESS}~B). 
 It is worth noting that the integrated density profile in Fig.~\ref{fig:NESS}~C and D displays a very small effect of the first local minimum (minor dip, captured only by time-course profiles at $t_1$ and $t_2$); this is reflected in relatively wide credible intervals for the inferred rate profile (grey ribbon in Fig.~\ref{fig:NESS}~B).
 On the other hand, the rate at the bottleneck is inferred
with very high confidence.
The covariance hyperparameters $l$ and $\sigma_f$  control how quickly the rate changes over $x$; these were slightly misestimated to 6.86~(95\%CI 4.63--7.16) and 
0.76~(95\%CI 0.75--0.82), respectively,
thus suggesting that increased wobbling in the rate profile is tolerated; minor patterns in the rate profile are in fact smoothed out and are essentially not identifiable in the density profiles obtained by integration (see Fig.~\ref{fig:posterior_zoomin}). The difficulty of sampling covariance hyperparameters is also addressed, e.g., in~\cite{Tegner2019}.
The predicted transient density profiles at $t=30,60,300$ also 
are in very good agreement with the input data 
(Fig.~\ref{fig:NESS}~D;  in fact, all sampled rate profiles yield similar time-course density profiles despite wide CIs in certain regions (see also Fig.~\ref{fig:posterior_zoomin}).

Applying this method to real-world data requires setting the
value of $\tilde{p}_{\mathrm{max}}$ to an upper limit of prior expectations on the
elongation rate. As this has been estimated at around $2\times 10^3$ Nt/min in previous studies~\cite{Jonkers2014},
we set $\tilde{p}_{\mathrm{max}}=6\times 10^3$ Nt/min as an arguably safe upper bound.
Literature results can be also used to set bounds on the prior for $\kappa$,
which regularises the estimation problem~\cite{Rasmussen2006}. From cultured human cell lines, the total number $\Rho$ of bound PolII molecules per cell is estimated to be between $\Rho_{\mathrm{min}}=11\times 10^5$ and $\Rho_{\mathrm{max}}=18\times 10^5$~\cite{Kimura1999}. This is related to the total number $Y$ of ChIP-seq counts by $\Rho=\kappa \, Y$. Based on these heuristic considerations, we set $\kappa_{\mathrm{min}}=\Rho_{\mathrm{min}} / Y$ and $\kappa_{\mathrm{max}}=\Rho_{\mathrm{max}} / Y$.
All remaining hyperparameters were set identical to the previous simulation experiment.

\begin{figure*}
    \includegraphics[width=0.90\linewidth]{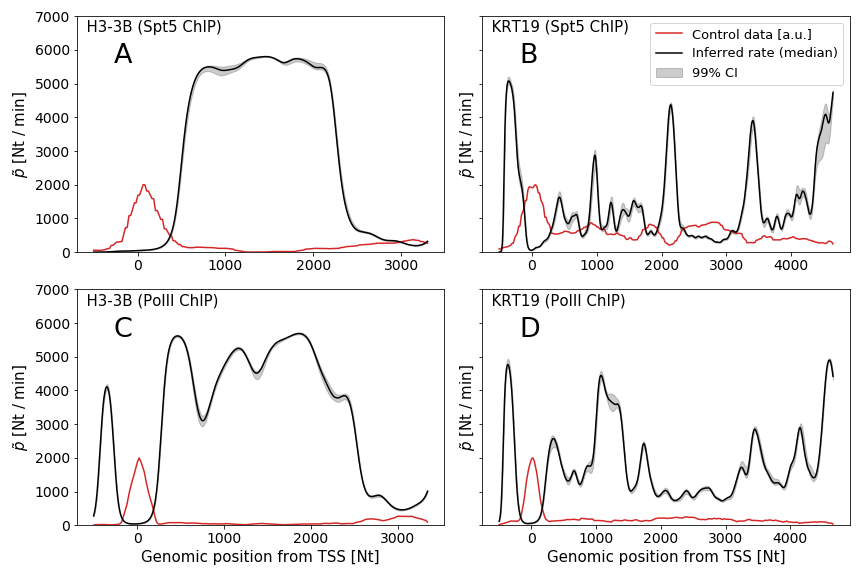}
    \caption{\label{fig:rate_profiles}
    Inferred rate profiles from Spt5 ChIP-seq (A-B) and PolII ChIP-seq (C-D) for genes H3-B3 (A-C) and KRT19 (B-D) (black lines are posterior medians, shaded area are 99\% credible intervals), with the latter gene showing a distinctive jagged profiles. Both genes show the lowest rates in proximity of the transcription starting site (TSS). Red lines are unperturbed ChIP-seq signals in arbitrary units (a.u.).
    }
\end{figure*}

\begin{figure}
    \includegraphics[width=0.5\linewidth]{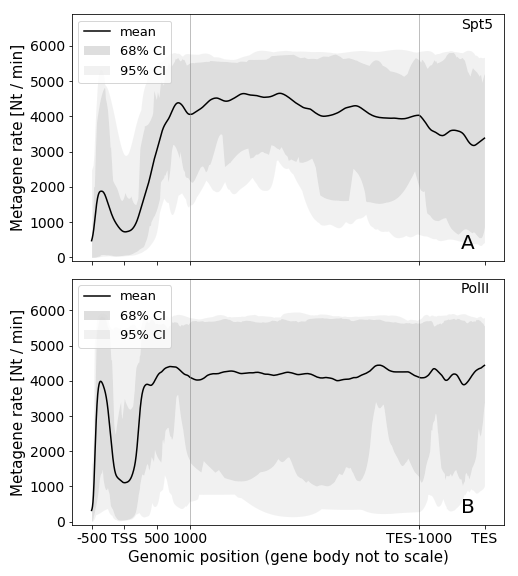}    
    \caption{
    \label{fig:metagenes}
    Metagene rates from PolII ChIP-seq (A) and Spt5 ChIP-seq  (B) data.
    By construction, the metagene analysis conserves the length scale only in proximity of TSS and TES. Upstream of TSS the Spt5 ChIP-seq yields lower average rate than PolII ChIP-seq, as this assay does not detect PolII poised to move downstream.}
\end{figure}

\begin{figure*}
    \includegraphics[width=0.90\linewidth]{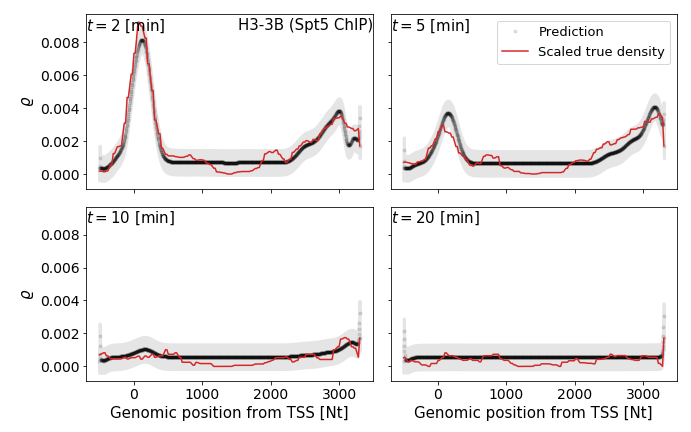}
    \caption{
         \label{fig:pos_predict}
    Predicted density profiles for gene H3-3B from Spt5 ChIP-seq $2$, $5$, $10$, and $20$ minutes after Trp treatment as samples from posterior predictive distribution. Grey area is 95\% credible interval due to noise model.
    }
\end{figure*}

The results from different genes show  a variety of rate
profiles which share similar patterns
(Fig.~\ref{fig:rate_profiles}).
The most important observation  is that, in all genes considered, the rates vary strongly with the genomic position,
with local minima corresponding to regions where PolIIs 
slow down or pause.
In order to look for average patterns, it is desirable to aggregate data from all genes.
As genes have different lengths (which in our sample range from 16,680 to 59,880 Nts), we stretch all the rate profiles in the region from TSS+1000 to TES-1000 Nts to the same support length and then average over the genes at each position. This yields the summaries illustrated in Fig.~\ref{fig:metagenes} which we refer to as \textit{metagene} rates and are akin to
the so-called metagene profiles~\cite{JolyBeauparlant2016}.
Rates are typically lower near the TSS than in the gene body, where elongation approaches its highest rate.
The behaviour in proximity of the TES is less definite, with rates varying several fold among the different genes.
At the TSS the rate typically dips down consistently with the presence of  strong  and widespread pausing in this region.
Further downstream in the gene body the rate increases to its highest average value.
While the dip is evident in both Spt5 and PolII results,
it is worth noting that upstream the TSS
the average rate inferred from PolII data is higher than that from Spt5.
We argue that this difference is due to the fact the former also include
poised PolIIs which are not strongly bound to the template and can quickly move towards the TSS before being engaged in transcription.
A by-product of the fitting procedure is the estimate of the occupation density $\varrho(x,t)=\kappa \, y(x,t)$, as illustrated in Fig.~\ref{fig:NESS}~D for the simulation experiment and
Figs.~\ref{fig:pos_predict},
\ref{fig:s1_pos_predict_}, \ref{fig:s2_pos_predict_}, and \ref{fig:s3_pos_predict_} for selected genes.
The predicted densities are typically very low (total predicted number of PolIIs in a gene is in the order of $10^{-1}$), thus suggesting that crowding and congestion of PolIIs into a gene might not be substantial even proximal to rate minima.

\section{Discussion}
\label{sec:discussion}
We developed a general Bayesian framework to study  the dynamics of a one-dimensional transport model given time-resolved density profiles. The general problem addressed here is the identification of the PDE parameters that best describe data as a subset of the true PDE solution (see, e.g., \cite{Rudy2017, Raissi2019, Berg2019, Tegner2019} and references therein).
We focused on the hydrodynamic TASEP with smoothly-varying 
jump rates (which are the parameters to be inferred) as a paradigmatic and well-characterised model of transport.
By means of its application to ChIP-seq time-course data, 
we inferred the rate of PolII elongation  as a function of the genomic position in selected genes.
This rate is not constant but varies within the gene body. It typically dips down nearby the TSS, confirming widespread pausing in this region, while in the bulk the rate also varies between genes.
Low predicted densities suggest that the pausing did not cause congestion or crowding.  This is an important observation, as factor crowding has been experimentally observed and associated with regulated gene expression in  synthetic and mammalian cell systems~\cite{Tan2013,Hnisz2017,Plys2018,Boija2018, Cavallaro2019}. Our analysis supports the view that this phenomenon does not happen between PolIIs bound to the gene but likely occurs in suspension in the nucleoplasm, as described,
e.g., in~\cite{Papantonis2013, Boehning2018, Plys2018, Cavallaro2019, Cramer2019, Wei2020}.

The inference here is complicated by the high dimensionality of the parameter space (which grows as the genes' length increases). We addressed this by assuming a Gaussian process latent prior for the jump-rate profile and using elliptic slice sampling as an appropriate MCMC algorithm.
 The sampling requires multiple evaluations of the likelihood of equation~\eqref{eq:loglikelihood}. This in turns requires numerically integrating equation~\eqref{eq:non_linear_PDE}, which is also slower in longer genes (require larger integration grids, see Appendix \ref{app:Numerical_integration}).

This study of molecular dynamics is also subject to  limitations.
While ChIP-seq is a widely-used assay to quantify the abundances of DNA-bound PolII, studies suggest that  it has limited resolution (between 150-300 Nts)  and might be subject to technical issues~\cite{Park2009}.
 Most importantly ChIP-seq profiles are obtained from the aggregation of sequencing reads from many cells, which hides variation within the cell population.
The transcription of mRNA is a very complex process and it may be interesting to include
features not encoded in the model used for this study.
Other  TASEP variants, such as those incorporating non-Markovian jump dynamics~\cite{Khoromskaia2014, Concannon2014} or Langmuir kinetics~\cite{Parmeggiani2004}, are relevant for the modelling of PolII recycling and its early detachment from DNA~\cite{Steurer2018,Cavallaro2019}.
An assumption of the TASEP is that particles stay in a site for an exponentially distributed waiting time. Variants of TASEP in which defects appear and disappear randomly on any site (and thus slow down the movement of particles or even block it completely)  have been introduced in physics literature and can account for occasionally long pausing times \cite{Wang2014, Waclaw2019} (see also \cite{Concannon2014, Khoromskaia2014}), with defect dynamics representing the effects of pausing and elongation factors.
Modelling advancements that combine
the site-specific pausing  with long pausing times and extended particle size, supplemented by an appropriate inference scheme such as the one presented here, would be  an important additional potential area for research and application.
Potential extensions of our work also include estimation of the parameters
that encode the system's size and asymmetry ($a$ and $b$, respectively) and the
boundary values.
Statistical mechanics literature is rich in quantitative 
studies of TASEPs with particles that occupy more than one lattice site~\cite{Shaw2003, Schonherr2004, Schonherr2005},
some generalized to include site-dependent elongation rates
or localised defects~\cite{Shaw2004, Shaw2004a, Dong2007, Erdmann-Pham2019}. These studies have been used to describe protein translation and could be useful to predict PolII-size effects in gene expression, although, 
 with genes much longer than PolIIs, ChIP-seq limited resolution, and very low PolII coverage density, the observable correction would arguably be minor.
In fact, including more features plausibly requires sequencing assays of higher resolution than ChIP-seq and comes at the cost of increased computational burden and decreased tractability.
Conversely, the chosen TASEP with smoothly-varying 
jump rates is simple and yet is able to reveal PolII elongation slowing down and speeding up at certain genomic locations.
Due to its generality, our approach also serves as a template for future studies seeking to shed light on complex transport phenomena.

\section*{Materials and Methods}
Spt5 and PolII ChIP-seq data mapped to the hg19 University of California at Santa Cruz human genome  were downloaded from Gene Expression Omnibus
(GEO, \url{http://www.ncbi.nlm.nih.gov/geo}),
accession number GSE117006.
We filtered the list of genes from the reference genome to only contain those with unique gene symbols on chromosomes 1–22 and X, thus excluding alternatively spliced genes.
Hg19 gene coordinates were flanked 500 Nts upstream the TSS in order to include poised PolII.
The 20 non-overlapping genes with the highest coverage of Spt5 ChIP-seq reads were selected.
All simulation codes are written in   c++ and  Python (v3.7.1),
with the PDE solver using Numba JIT compiler (v0.41.0)~\cite{Lam2015}
(\url{https://github.com/mcavallaro/dTASEP-fit}).

\section*{Acknowledgments}
This research utilised WISB computational facilities (grant ref: BB/M017982/1) funded under the UK Research Councils' Synthetic Biology for Growth programme.
MC acknowledges support from Matt J.~Keeling and Health Data Research UK, which is funded by the UK Medical Research Council, EPSRC, Economic and Social Research Council, Department of Health and Social Care (England), Chief Scientist Office of the Scottish Government Health and Social Care Directorates, Health and Social Care Research and Development Division (Welsh Government), Public Health Agency (Northern Ireland), British Heart Foundation and the Wellcome Trust. DH is funded by EPSRC (grant no.~EP/T002794/1). RD is funded by EPSRC (grant nos.~EP/V025899/1, EP/T017112/1) and NERC (grant no.~NE/T00973X/1). 
We thank Carlo Albert and Jie Zhang for valuable comments
and the Warwick Bioinformatics RTP for sharing computational resources.

\section*{Author contributions}
Conceptualization  MC  DH  RD;  Data  curation  MC DH; Formal analysis MC RD; Investigation MC; Software MC YW RD; Writing -- original draft MC; Writing -- review \& editing MC DH RD.

\bibliographystyle{vancouver.bst}
\bibliography{revtex}

\begin{thebibliography}{10}

\bibitem{Munsky2012}
Munsky B, Neuert G, van Oudenaarden A.
\newblock {Using Gene Expression Noise to Understand Gene Regulation}.
\newblock Science. 2012 apr;336(6078):183-7.
\newblock doi:10.1126/science.1216379.

\bibitem{Rajala2010}
Rajala T, H\"akkinen A, Healy S, Yli-Harja O, Ribeiro AS.
\newblock Effects of Transcriptional Pausing on Gene Expression Dynamics.
\newblock PLoS Comput Biol. 2010 03;6(3):1-12.
\newblock doi:10.1371/journal.pcbi.1000704.

\bibitem{Jonkers2015}
Jonkers I, Lis JT.
\newblock {Getting up to speed with transcription elongation by RNA polymerase
  II}.
\newblock Nat Rev Mol Cell Biol. 2015 mar;16(3):167-77.
\newblock doi:10.1038/nrm3953.

\bibitem{Mayer2017}
Mayer A, Landry HM, Churchman LS.
\newblock {Pause \& go: from the discovery of RNA polymerase pausing to its
  functional implications}.
\newblock Curr Opin Cell Biol. 2017 jun;46:72-80.
\newblock doi:10.1016/j.ceb.2017.03.002.

\bibitem{Adelman2012}
Adelman K, Lis JT.
\newblock {Promoter-proximal pausing of RNA polymerase II: emerging roles in
  metazoans.}
\newblock Nat Rev Genet. 2012 sep;13(10):720-31.
\newblock doi:10.1038/nrg3293.

\bibitem{Liu2015}
Liu X, Kraus WL, Bai X.
\newblock {Ready, pause, go: Regulation of RNA polymerase II pausing and
  release by cellular signaling pathways}.
\newblock Trends Biochem Sci. 2015 sep;40(9):516-25.
\newblock doi:10.1016/j.tibs.2015.07.003.

\bibitem{peccoud1995markovian}
Peccoud J, Ycart B.
\newblock Markovian modeling of gene-product synthesis.
\newblock Theor Popul Biol. 1995;48(2):222-34.

\bibitem{Tripathi2008}
Tripathi T, Chowdhury D.
\newblock {Interacting RNA polymerase motors on a DNA track: Effects of traffic
  congestion and intrinsic noise on RNA synthesis}.
\newblock Phys Rev E. 2008 jan;77(1):011921.
\newblock doi:10.1103/PhysRevE.77.011921.

\bibitem{Dobrzynski2009}
Dobrzynski M, Bruggeman FJ.
\newblock {Elongation dynamics shape bursty transcription and translation}.
\newblock Proc Natl Acad Sci. 2009 feb;106(8):2583-8.
\newblock doi:10.1073/pnas.0803507106.

\bibitem{Cao2020}
Cao Z, Filatova T, Oyarzún DA, Grima R.
\newblock A Stochastic Model of Gene Expression with Polymerase Recruitment and
  Pause Release.
\newblock Biophys J. 2020;119(5):1002-14.
\newblock doi:10.1016/j.bpj.2020.07.020.

\bibitem{Szavits-Nossan2022}
Szavits-Nossan J, Grima R.
\newblock Steady-state distributions of nascent RNA for general initiation
  mechanisms.
\newblock Phys Rev Res. 2023 Jan;5:013064.
\newblock doi:10.1103/PhysRevResearch.5.013064.

\bibitem{MacDonald1968}
MacDonald CT, Gibbs JH, Pipkin AC.
\newblock {Kinetics of biopolymerization on nucleic acid templates}.
\newblock Biopolymers. 1968 jan;6(1):1-25.
\newblock doi:10.1002/bip.1968.360060102.

\bibitem{MacDonald69}
MacDonald CT, Gibbs JH.
\newblock Concerning the kinetics of polypeptide synthesis on polyribosomes.
\newblock Biopolymers. 1969;7(5):707-25.
\newblock doi:https://doi.org/10.1002/bip.1969.360070508.

\bibitem{Spitzer1970}
Spitzer F.
\newblock {Interaction of Markov processes}.
\newblock Adv Math. 1970 oct;5(2):246-90.
\newblock doi:10.1016/0001-8708(70)90034-4.

\bibitem{Benassi1987}
Benassi A, Fouque JP.
\newblock {Hydrodynamical Limit for the Asymmetric Simple Exclusion Process}.
\newblock Ann Probab. 1987 apr;15(2):546-60.
\newblock doi:10.1214/aop/1176992158.

\bibitem{Stinchcombe2011}
Stinchcombe RB, {De Queiroz} SLA.
\newblock {Smoothly varying hopping rates in driven flow with exclusion}.
\newblock Phys Rev E. 2011;83(6):1-12.
\newblock doi:10.1103/PhysRevE.83.061113.

\bibitem{Lakatos2006}
Lakatos G, O'Brien J, Chou T.
\newblock {Hydrodynamic mean-field solutions of 1D exclusion processes with
  spatially varying hopping rates}.
\newblock J Phys A: Math Gen. 2006;39(10):2253-64.

\bibitem{Harris2004}
Harris RJ, Stinchcombe RB.
\newblock {Disordered asymmetric simple exclusion process: Mean-field
  treatment}.
\newblock Phys Rev E. 2004 jul;70(1):016108.
\newblock doi:10.1103/PhysRevE.70.016108.

\bibitem{Rasmussen2006}
Rasmussen CE, Williams CKI.
\newblock {Gaussian processes for machine learning}.
\newblock MIT Press; 2006.

\bibitem{WaMaina2014}
wa~Maina C, Honkela A, Matarese F, Grote K, Stunnenberg HG, Reid G, et~al.
\newblock {Inference of RNA Polymerase II Transcription Dynamics from Chromatin
  Immunoprecipitation Time Course Data}.
\newblock PLoS Comput Biol. 2014 may;10(5):1-17.
\newblock doi:10.1371/journal.pcbi.1003598.

\bibitem{Honkela2015}
Honkela A, Peltonen J, Topa H, Charapitsa I, Matarese F, Grote K, et~al.
\newblock {Genome-wide modeling of transcription kinetics reveals patterns of
  RNA production delays}.
\newblock Proc Natl Acad Sci. 2015 oct;112(42):13115-20.
\newblock doi:10.1073/pnas.1420404112.

\bibitem{Erickson2018}
Erickson B, Sheridan RM, Cortazar M, Bentley DL.
\newblock {Dynamic turnover of paused pol II complexes at human promoters}.
\newblock Genes Dev. 2018 sep;32(17-18):1215-25.
\newblock doi:10.1101/gad.316810.118.

\bibitem{Park2009}
Park PJ.
\newblock {ChIP-seq: Advantages and challenges of a maturing technology}.
\newblock Nat Rev Genet. 2009;10(10):669-80.
\newblock doi:10.1038/nrg2641.

\bibitem{Ehara2017}
Ehara H, Yokoyama T, Shigematsu H, Yokoyama S, Shirouzu M, Sekine Si.
\newblock {Structure of the complete elongation complex of RNA polymerase II
  with basal factors}.
\newblock Science. 2017 sep;357(6354):921-4.
\newblock doi:10.1126/science.aan8552.

\bibitem{Hu2015}
Hu B, Petela N, Kurze A, Chan KL, Chapard C, Nasmyth K.
\newblock {Biological chromodynamics: a general method for measuring protein
  occupancy across the genome by calibrating ChIP-seq}.
\newblock Nucleic Acids Res. 2015 jun:gkv670.
\newblock doi:10.1093/nar/gkv670.

\bibitem{Nojima2015}
Nojima T, Gomes T, Grosso ARF, Kimura H, Dye MJ, Dhir S, et~al.
\newblock {Mammalian NET-seq reveals genome-wide nascent transcription coupled
  to RNA processing}.
\newblock Cell. 2015;161(3):526-40.
\newblock doi:10.1016/j.cell.2015.03.027.

\bibitem{Core2008}
Core LJ, Waterfall JJ, Lis JT.
\newblock {Nascent RNA Sequencing Reveals Widespread Pausing and Divergent
  Initiation at Human Promoters}.
\newblock Science. 2008 dec;322(5909):1845-8.
\newblock doi:10.1126/science.1162228.

\bibitem{Kwak2013}
Kwak H, Fuda NJ, Core LJ, Lis JT.
\newblock {Precise Maps of RNA Polymerase Reveal How Promoters Direct
  Initiation and Pausing}.
\newblock Science. 2013 feb;339(6122):950.
\newblock doi:10.1126/science.1229386.

\bibitem{Zhang2021}
Zhang J, Cavallaro M, Hebenstreit D.
\newblock {Timing RNA polymerase pausing with TV-PRO-seq}.
\newblock Cell Rep Methods. 2021 sep;1(0):100083.
\newblock doi:10.1016/j.crmeth.2021.100083.

\bibitem{Jonkers2014}
Jonkers I, Kwak H, Lis JT.
\newblock {Genome-wide dynamics of Pol II elongation and its interplay with
  promoter proximal pausing, chromatin, and exons}.
\newblock eLife. 2014 apr;2014(3):1-25.
\newblock doi:10.7554/eLife.02407.

\bibitem{Chou2011}
Chou T, Mallick K, Zia RKP.
\newblock {Non-equilibrium statistical mechanics: from a paradigmatic model to
  biological transport}.
\newblock Rep Prog Phys. 2011 nov;74(11):116601.
\newblock doi:10.1088/0034-4885/74/11/116601.

\bibitem{Klumpp2008}
Klumpp S, Hwa T.
\newblock {Stochasticity and traffic jams in the transcription of ribosomal
  RNA: Intriguing role of termination and antitermination}.
\newblock Proc Natl Acad Sci. 2008 nov;105(47):18159-64.
\newblock doi:10.1073/PNAS.0806084105.

\bibitem{Cholewa-Waclaw2019}
Cholewa-Waclaw J, Shah R, Webb S, Chhatbar K, Ramsahoye B, Pusch O, et~al.
\newblock {Quantitative modelling predicts the impact of DNA methylation on RNA
  polymerase II traffic.}
\newblock Proc Natl Acad Sci. 2019 jul;116(30):14995-5000.
\newblock doi:10.1073/pnas.1903549116.

\bibitem{Tripathi2009}
Tripathi T, Sch{\"{u}}tz GM, Chowdhury D.
\newblock {RNA polymerase motors: dwell time distribution, velocity and
  dynamical phases}.
\newblock J Stat Mech Theory Exp. 2009 aug;2009(08):P08018.
\newblock doi:10.1088/1742-5468/2009/08/P08018.

\bibitem{Zia2011}
Zia RKP, Dong JJ, Schmittmann B.
\newblock {Modeling Translation in Protein Synthesis with TASEP: A Tutorial and
  Recent Developments}.
\newblock J Stat Phys. 2011 apr;144(2):405-28.
\newblock doi:10.1007/s10955-011-0183-1.

\bibitem{Szavits-Nossan2018}
Szavits-Nossan J, Ciandrini L, Romano MC.
\newblock {Deciphering mRNA Sequence Determinants of Protein Production Rate}.
\newblock Phys Rev Lett. 2018 mar;120(12):128101.
\newblock doi:10.1103/PhysRevLett.120.128101.

\bibitem{Erdmann-Pham2019}
Erdmann-Pham DD, {Dao Duc} K, Song YS.
\newblock The Key Parameters that Govern Translation Efficiency.
\newblock Cell Syst. 2020;10(2):183-92.e6.
\newblock doi:10.1016/j.cels.2019.12.003.

\bibitem{Lipowsky2001}
Lipowsky R, Klumpp S, Nieuwenhuizen TM.
\newblock {Random Walks of Cytoskeletal Motors in Open and Closed
  Compartments}.
\newblock Phys Rev Lett. 2001 aug;87(10):108101.
\newblock doi:10.1103/PhysRevLett.87.108101.

\bibitem{Chowdhury2000}
Chowdhury D, Santen L, Schadschneider A.
\newblock {Statistical physics of vehicular traffic and some related systems}.
\newblock Physics Reports. 2000 may;329(4-6):199-329.
\newblock doi:10.1016/S0370-1573(99)00117-9.

\bibitem{Wang2014}
Wang J, Pfeuty B, Thommen Q, Romano MC, Lefranc M.
\newblock {Minimal model of transcriptional elongation processes with pauses}.
\newblock Phys Rev E. 2014 nov;90(5):050701.
\newblock doi:10.1103/PhysRevE.90.050701.

\bibitem{Waclaw2019}
Waclaw B, Cholewa-Waclaw J, Greulich P.
\newblock {Totally asymmetric exclusion process with site-wise dynamic
  disorder}.
\newblock J Phys A: Math Theor. 2019 feb;52(6):065002.
\newblock doi:10.1088/1751-8121/aafb8a.

\bibitem{Kardar1986}
Kardar M, Parisi G, Zhang YC.
\newblock {Dynamic scaling of growing interfaces}.
\newblock Phys Rev Lett. 1986 mar;56(9):889-92.
\newblock doi:10.1103/PhysRevLett.56.889.

\bibitem{Bertini1997}
Bertini L, Giacomin G.
\newblock {Stochastic Burgers and KPZ equations from particle systems}.
\newblock Commun Math Phys. 1997;183(3):571-607.
\newblock doi:10.1007/s002200050044.

\bibitem{Derrida1993}
Derrida B, Evans MR, Hakim V, Pasquier V.
\newblock {Exact solution of a 1D asymmetric exclusion model using a matrix
  formulation}.
\newblock J Phys A: Math Gen. 1993 apr;26(7):1493-517.
\newblock doi:10.1088/0305-4470/26/7/011.

\bibitem{Lazarescu2015}
Lazarescu A.
\newblock {The physicist's companion to current fluctuations: one-dimensional
  bulk-driven lattice gases}.
\newblock J Phys A: Math Theor. 2015 dec;48(50):503001.

\bibitem{Hopf1950}
Hopf E.
\newblock {The partial differential equation ut + uux = $\mu$xx}.
\newblock Commun Pure Appl Math. 1950 sep;3(3):201-30.
\newblock doi:10.1002/cpa.3160030302.

\bibitem{Cole1951}
Cole JD.
\newblock {On a quasi-linear parabolic equation occurring in aerodynamics}.
\newblock Q Appl Math. 1951 oct;9(3):225-36.
\newblock doi:10.1090/qam/42889.

\bibitem{Nudler1997}
Nudler E, Mustaev A, Goldfarb A, Lukhtanov E.
\newblock {The RNA--DNA Hybrid Maintains the Register of Transcription by
  Preventing Backtracking of RNA Polymerase}.
\newblock Cell. 1997 apr;89(1):33-41.
\newblock doi:10.1016/S0092-8674(00)80180-4.

\bibitem{Julicher1998}
J{\"{u}}licher F, Bruinsma R.
\newblock {Motion of RNA polymerase along DNA: A stochastic model}.
\newblock Biophys J. 1998;74(3):1169-85.
\newblock doi:10.1016/S0006-3495(98)77833-6.

\bibitem{Wang2009}
Wang D, Bushnell DA, Huang X, Westover KD, Levitt M, Kornberg RD.
\newblock {Structural Basis of Transcription: Backtracked RNA Polymerase II at
  3.4 Angstrom Resolution}.
\newblock Science. 2009 may;324(5931):1203-6.
\newblock doi:10.1126/science.1168729.

\bibitem{gelman2013bayesian}
Gelman A, Carlin JB, Stern HS, Dunson DB, Vehtari A, Rubin DB.
\newblock Bayesian data analysis.
\newblock CRC press; 2013.

\bibitem{Tegner2019}
Tegn{\'e}r M, Roberts S.
\newblock A probabilistic approach to nonparametric local volatility.
\newblock arXiv preprint arXiv:190106021. 2019.

\bibitem{Murray2010}
Murray I, Adams R, MacKay D.
\newblock Elliptical slice sampling.
\newblock In: Proceedings of the thirteenth international conference on
  artificial intelligence and statistics. JMLR Workshop and Conference
  Proceedings; 2010. p. 541-8.

\bibitem{Kimura1999}
Kimura H, Tao Y, Roeder RG, Cook PR.
\newblock Quantitation of {RNA} polymerase {II} and its transcription factors
  in an {HeLa} cell: little soluble holoenzyme but significant amounts of
  polymerases attached to the nuclear substructure.
\newblock Mol Cell Biol. 1999 aug;19(8):5383-92.
\newblock doi:10.1128/mcb.19.8.5383.

\bibitem{JolyBeauparlant2016}
{Joly Beauparlant} C, Lamaze FC, Desch{\^{e}}nes A, Samb R, Lema{\c{c}}on A,
  Belleau P, et~al.
\newblock {metagene Profiles Analyses Reveal Regulatory Element's
  Factor-Specific Recruitment Patterns}.
\newblock PLoS Comput Biol. 2016 aug;12(8):1-12.
\newblock doi:10.1371/journal.pcbi.1004751.

\bibitem{Rudy2017}
Rudy SH, Brunton SL, Proctor JL, Kutz JN.
\newblock {Data-driven discovery of partial differential equations}.
\newblock Sci Adv. 2017 apr;3(4):e1602614.
\newblock doi:10.1126/sciadv.1602614.

\bibitem{Raissi2019}
Raissi M, Perdikaris P, Karniadakis GE.
\newblock {Physics-informed neural networks: A deep learning framework for
  solving forward and inverse problems involving nonlinear partial differential
  equations}.
\newblock Journal of Computational Physics. 2019;378:686-707.
\newblock doi:10.1016/J.JCP.2018.10.045.

\bibitem{Berg2019}
Berg J, Nystr{\"{o}}m K.
\newblock {Data-driven discovery of PDEs in complex datasets}.
\newblock Journal of Computational Physics. 2019 may;384:239-52.
\newblock doi:10.1016/j.jcp.2019.01.036.

\bibitem{Tan2013}
Tan C, Saurabh S, Bruchez MP, Schwartz R, Leduc P.
\newblock {Molecular crowding shapes gene expression in synthetic cellular
  nanosystems}.
\newblock Nat Nanotechnol. 2013 jul;8(8):602-8.
\newblock doi:10.1038/nnano.2013.132.

\bibitem{Hnisz2017}
Hnisz D, Shrinivas K, Young RA, Chakraborty AK, Sharp PA.
\newblock {A Phase Separation Model for Transcriptional Control}.
\newblock Cell. 2017 mar;169(1):13-23.
\newblock doi:10.1016/j.cell.2017.02.007.

\bibitem{Plys2018}
Plys AJ, Kingston RE.
\newblock {Dynamic condensates activate transcription}.
\newblock Science. 2018;361(6400):329-30.
\newblock doi:10.1126/science.aau4795.

\bibitem{Boija2018}
Boija A, Klein IA, Sabari BR, Dall'Agnese A, Coffey EL, Zamudio AV, et~al.
\newblock {Transcription Factors Activate Genes through the Phase-Separation
  Capacity of Their Activation Domains}.
\newblock Cell. 2018 dec;175(7):1842-55.e16.
\newblock doi:10.1016/j.cell.2018.10.042.

\bibitem{Cavallaro2019}
Cavallaro M, Walsh MD, Jones M, Teahan J, Tiberi S, Finkenst{\"{a}}dt B, et~al.
\newblock 3'-5' crosstalk contributes to transcriptional bursting.
\newblock Genome Biol. 2021 dec;22(1):56.
\newblock doi:10.1186/s13059-020-02227-5.

\bibitem{Papantonis2013}
Papantonis A, Cook PR.
\newblock {Transcription factories: Genome organization and gene regulation}.
\newblock Chemical Reviews. 2013 nov;113(11):8683-705.
\newblock doi:10.1021/cr300513p.

\bibitem{Boehning2018}
Boehning M, Dugast-Darzacq C, Rankovic M, Hansen AS, Yu T, Marie-Nelly H,
  et~al.
\newblock {RNA polymerase II clustering through carboxy-terminal domain phase
  separation}.
\newblock Nature Structural and Molecular Biology. 2018 sep;25(9):833-40.
\newblock doi:10.1038/s41594-018-0112-y.

\bibitem{Cramer2019}
Cramer P.
\newblock {Organization and regulation of gene transcription}.
\newblock Nature. 2019.
\newblock doi:10.1038/s41586-019-1517-4.

\bibitem{Wei2020}
Wei MT, Chang YC, Shimobayashi SF, Shin Y, Strom AR, Brangwynne CP.
\newblock {Nucleated transcriptional condensates amplify gene expression}.
\newblock Nature Cell Biology. 2020.
\newblock doi:10.1038/s41556-020-00578-6.

\bibitem{Khoromskaia2014}
Khoromskaia D, Harris RJ, Grosskinsky S.
\newblock {Dynamics of non-Markovian exclusion processes}.
\newblock J Stat Mech Theory Exp. 2014 dec;2014(12):P12013.
\newblock doi:10.1088/1742-5468/2014/12/P12013.

\bibitem{Concannon2014}
Concannon RJ, Blythe RA.
\newblock {Spatiotemporally Complete Condensation in a Non-Poissonian Exclusion
  Process}.
\newblock Phys Rev Lett. 2014 feb;112(5):050603.
\newblock doi:10.1103/PhysRevLett.112.050603.

\bibitem{Parmeggiani2004}
Parmeggiani A, Franosch T, Frey E.
\newblock {Totally asymmetric simple exclusion process with Langmuir kinetics}.
\newblock Phys Rev E. 2004 oct;70(4):046101.
\newblock doi:10.1103/PhysRevE.70.046101.

\bibitem{Steurer2018}
Steurer B, Janssens RC, Geverts B, Geijer ME, Wienholz F, Theil AF, et~al.
\newblock {Live-cell analysis of endogenous GFP-RPB1 uncovers rapid turnover of
  initiating and promoter-paused RNA Polymerase II.}
\newblock Proc Natl Acad Sci. 2018 may;115(19):E4368-76.
\newblock doi:10.1073/pnas.1717920115.

\bibitem{Shaw2003}
Shaw LB, Zia RKP, Lee KH.
\newblock {Totally asymmetric exclusion process with extended objects: A model
  for protein synthesis}.
\newblock Phys Rev E. 2003 aug;68(2):021910.
\newblock doi:10.1103/PhysRevE.68.021910.

\bibitem{Schonherr2004}
Sch{\"{o}}nherr G, Sch{\"{u}}tz GM.
\newblock {Exclusion process for particles of arbitrary extension: hydrodynamic
  limit and algebraic properties}.
\newblock J phys A: Math Gen. 2004 aug;37(34):8215-31.
\newblock doi:10.1088/0305-4470/37/34/002.

\bibitem{Schonherr2005}
Sch{\"{o}}nherr G.
\newblock {Hard rod gas with long-range interactions: Exact predictions for
  hydrodynamic properties of continuum systems from discrete models}.
\newblock Phys Rev E. 2005 feb;71(2):026122.
\newblock doi:10.1103/PhysRevE.71.026122.

\bibitem{Shaw2004}
Shaw LB, Sethna JP, Lee KH.
\newblock {Mean-field approaches to the totally asymmetric exclusion process
  with quenched disorder and large particles}.
\newblock Phys Rev E. 2004 aug;70(2):021901.
\newblock doi:10.1103/PhysRevE.70.021901.

\bibitem{Shaw2004a}
Shaw LB, Kolomeisky AB, Lee KH.
\newblock {Local inhomogeneity in asymmetric simple exclusion processes with
  extended objects}.
\newblock J Phys A: Math Gen. 2004 feb;37(6):2105-13.
\newblock doi:10.1088/0305-4470/37/6/010.

\bibitem{Dong2007}
Dong JJ, Schmittmann B, Zia RKP.
\newblock {Inhomogeneous exclusion processes with extended objects: The effect
  of defect locations}.
\newblock Phys Rev E. 2007 nov;76(5):051113.
\newblock doi:10.1103/PhysRevE.76.051113.

\bibitem{Lam2015}
Lam SK, Pitrou A, Seibert S.
\newblock {Numba: a LLVM-based Python JIT compiler}.
\newblock In: Proceedings of the Second Workshop on the LLVM Compiler
  Infrastructure in HPC - LLVM '15. New York, New York, USA: ACM Press; 2015.
  p. 1-6.
\newblock doi:10.1145/2833157.2833162.

\end{thebibliography}

\begin{widetext}

\appendix

\clearpage
\setcounter{figure}{0}
\setcounter{table}{0}
\renewcommand{\thefigure}{S\arabic{figure}}
\renewcommand{\thetable}{S\arabic{table}}

\section{Cole-Hopf transform}
\label{app:Cole-Hopf_transform}
Let us define a ``height'' variable $h(x,t)$ as follows:
\begin{equation}
  \varrho(x,t) =:\frac{\partial}{\partial x} h(x,t) +  \frac{1}{2}.
  \label{eq:height}
\end{equation}
Substituting this into equation~\eqref{eq:non_linear_PDE} of the main text yields
\begin{equation}
  \frac{\partial }{\partial t} \frac{\partial }{\partial x}  h(x,t) = - \frac{\partial}{\partial x} 
  \left\{ \lambda(x) \left( \frac{\partial}{\partial x} h(x,t) +  \frac{1}{2}\right) \left(\frac{1}{2} -  \frac{\partial}{\partial x} h(x,t) \right) -
   \nu(x) \frac{\partial^2}{\partial^2 x} h(x,t) \right\}
\end{equation}
and integrating over $x$ gives
\begin{equation}
  \frac{\partial }{\partial t} h(x,t) = -  
  \lambda(x) \left( \frac{\partial}{\partial x} h(x,t) +  \frac{1}{2}\right) \left(\frac{1}{2} -  \frac{\partial}{\partial x} h(x,t) \right) +
   \nu(x) \frac{\partial^2}{\partial^2 x} h(x,t) + f(t),
   \label{eq:KPZ}
\end{equation}
which is a  noiseless version of the paradigmatic growth model studied by Kardar, Parisi, and Zhang~\cite{Kardar1986}
up to an arbitrary integration function $f(t)$ constant in $x$.

This equation is linearised by means of the transformation
\begin{equation}
  h(x,t) =:  \frac{a}{2} \frac{1+b}{1-b} \ln u(x,t) + F(t),
  \label{eq:field}
\end{equation}
which obviously implies
\begin{equation}
  \frac{\partial}{\partial x}h(x,t) =  \frac{a}{2} \frac{1+b}{1-b} \frac{u_x(x,t)}{u(x,t)}.
\end{equation}
We refer to $u(x,t)$ as the ``field''.

Merely substituting~\eqref{eq:field} into equation~\eqref{eq:KPZ} and choosing $F(t)$
to remove the term $f(t)$ yields
\begin{equation}
  \frac{a}{2} \frac{1+b}{1-b} \frac{u_t(x,t)}{u(x,t)} = -  
  \left\{ \lambda(x) \left[ \frac{1}{4}  - \left(\frac{a}{2} \frac{1+b}{1-b} \frac{u_x(x,t)}{u(x,t)}\right)^2 \right]
   - \nu(x) \frac{a}{2} \frac{1+b}{1-b} \frac{\partial}{\partial x} \frac{u_x(x,t)}{u(x,t)} \right\},
\end{equation}

% \begin{equation}
\begin{multline}
  \frac{a}{2} \frac{1+b}{1-b} \frac{u_t(x,t)}{u(x,t)} = \\
  -  a \tilde{p}(x) 
  \left\{   \left[ \frac{(1-b)}{4} - \frac{a^2}{4} \frac{(1+b)^2}{(1-b)} \frac{(u_x(x,t))^2}{u^2(x,t)} \right]
   - \frac{a^2}{4} \frac{(1+b)^2}{1-b}  \left( \frac{u_{xx}(x,t)}{u(x,t)}  - \frac{(u_x(x,t))^2}{u^2(x,t)} \right) \right\},
% \end{equation}
\end{multline}

% \begin{equation}
\begin{multline}
\frac{1}{2} \frac{1+b}{1-b} \frac{u_t(x,t)}{u(x,t)} = \\ - \tilde{p}(x) 
  \left\{ \frac{(1-b)}{4} - \cancel{\frac{a^2}{4} \frac{(1+b)^2}{(1-b)} \frac{(u_x(x,t))^2}{u^2(x,t)}}
   - \frac{a^2}{4} \frac{(1+b)^2}{1-b}   \frac{u_{xx}(x,t)}{u(x,t)}  + \cancel{\frac{a^2}{4} \frac{(1+b)^2}{1-b} \frac{(u_x(x,t))^2}{u^2(x,t)}}  \right\},
% \end{equation}
\end{multline}

\begin{equation}
   \frac{\partial }{\partial t} u(x,t) =  \frac{\tilde{p}(x)} {2}
  \left\{{a^2} (1+b)   \frac{\partial^2}{\partial^2 x}u (x,t) - \frac{(1 - b)^2}{1 + b} u(x,t) \right\},
  \label{eq:linear_PDE_}
\end{equation}
which, for  the totally asymmetric case $b=0$, can be simplified to
\begin{equation}
   \frac{\partial }{\partial t} u(x,t) =  \frac{\tilde p(x)} {2}
  \left\{{a^2}  \frac{\partial^2}{\partial^2 x}u(x,t) - u(x,t) \right\}.
  \label{eq:linear}
\end{equation}
To the best of our knowledge, this parametrisation  was first derived in reference~\cite{Harris2004}.
If $\tilde{p}(x)=p$, it is possible to incorporate a term $-a\,p\,(1-b) \, t / 4$ in $F(t)$ to eliminate the second term on
the right-hand side of equations~\eqref{eq:linear_PDE_} and \eqref{eq:linear}, thus further simplifying these to diffusion equations.

Equation~\eqref{eq:linear_PDE_} preserves the full dynamics of equation~\eqref{eq:non_linear_PDE} of the main text and is linear,
thus it is easier to treat numerically than the latter.
In the next sections we
elucidate the numerical scheme for its integration for $b=0$.

We impose Dirichlet boundary conditions for $\varrho(x,t)$ at $x=0$ and $x=L$, i.e.,
\begin{equation}
\begin{aligned}
    \varrho(0,t) &= \varrho_{\mathrm{left}}(t), \\
    \varrho(L,t) &= \varrho_{\mathrm{right}}(t),\\
\end{aligned}
\end{equation}
$\forall t>0$, which implies Neumann conditions for the height $h(x,t)$:
\begin{equation}
\begin{aligned}
    h_x(x,t)|_{x=0} &= \varrho_{\mathrm{left}} (t)- \frac{1}{2}, \\
    h_x(x,t)|_{x=L} &= \varrho_{\mathrm{right}} (t)- \frac{1}{2}, 
\end{aligned}
\end{equation}
and Robin (mixed) boundary conditions for the field $u(x,t)$:
\begin{align}
    (2\varrho_{\mathrm{left}}(t) -1) u(0,t) / a  -u_x(x,t)|_{x=0} & = 0, \label{eq:R1} \\
    (2\varrho_{\mathrm{right}}(t) - 1) u(L,t) / a  -u_x(x,t)|_{x=L}  & = 0. \label{eq:R2}
\end{align}

\section{Numerical integration}
\label{app:Numerical_integration}
The solution $u^*$ of equation~\eqref{eq:linear} at the coordinates $(\mathsf{i}\,\Delta x, \mathsf{j}\,\Delta t)$
is approximated by $U_{\mathsf j,\mathsf i}$ ($\mathsf j=0,1,\ldots,T$
and $\mathsf i = 0,1,\ldots,\mathsf N$ with $\mathsf N=L/\Delta x$)
which is computed using the forward Euler explicit iterative procedure
\begin{equation}
    U_{\mathsf j+1,\mathsf i} = U_{\mathsf j, \mathsf i} + P_{\mathsf i} \,\left[ a^2 \frac{U_{\mathsf j, \mathsf i-1} - 2 U_{\mathsf j, \mathsf i} + U_{\mathsf j, \mathsf i+1}}{\Delta x^2} - U_{\mathsf j, \mathsf i} \right] \Delta t ,
    \label{eq:forward_euler}
\end{equation}
where $P_{\mathsf i}=\tilde{p}(\mathsf i\, \Delta x)$.
The approximated density $\rho_{\mathsf{j},\mathsf{i}} \approx \varrho(\mathsf{i} \Delta x, \mathsf{j} \Delta t)$ is recovered from $U_{\mathsf j, \mathsf i}$ by means of the transformation
\begin{equation}
\begin{aligned}
    H_{\mathsf j,\mathsf i} &= \frac{a}{2}  \, \ln U_{\mathsf j,\mathsf i},  \\
    \rho_{\mathsf j,\mathsf i} &= \frac{1}{2} + \frac{H_{\mathsf j, \mathsf i+1} - H_{\mathsf j, \mathsf  i-1}}{2 \Delta x}.
    \label{eq:inverse_discrete_cole_hopf}
\end{aligned}  
\end{equation}
 Initial values $U_{0,\mathsf i}$, $\mathsf i = 0,1,\ldots,\mathsf N$, are obtained by means of
the discrete Cole-Hopf transform
\begin{equation}
      U_{\mathsf j,\mathsf i} = \exp \left( \sum_{\mathsf k=0}^{\mathsf i} \frac{1}{a}(2  \rho_{\mathsf j, \mathsf k} - 1)  \Delta{x} \right),
      \label{eq:discrete_cole_hopf}
\end{equation}
at $\mathsf j=0$ and given $\rho_{0, \mathsf i}$.
We deal with the boundary conditions by introducing ghost grid points
$(\mathsf j,-1)$ and $(\mathsf  j,\mathsf N+1)$, with $\mathsf j=0,1,2,\ldots,T$), where the values of $U$ are determined according
to the  procedure detailed below.

We use a central difference formula to approximate the derivatives 
\begin{align}
    u_x|_{x=0}(x,t) &\approx \frac{U_{\mathsf j,1} - U_{\mathsf j,-1}}{2\Delta x},\\
    u_x|_{x=L}(x,t) &\approx \frac{U_{\mathsf j,\mathsf N+1} - U_{\mathsf j, \mathsf N-1}}{2 \Delta x},
\end{align}
while $u(0,t) \approx U_{\mathsf j,1}$ and $u(L,t) \approx U_{\mathsf j, \mathsf N}$.
Substituting into equations \eqref{eq:R1}--\eqref{eq:R2}:
\begin{align}
    U_{\mathsf j, -1} & \approx U_{\mathsf j,1} - 2\Delta x [2 \, \rho_{\mathsf j, 0 } - 1] \frac{1}{a} U_{\mathsf j,0},\\
    U_{\mathsf j,\mathsf N+1} & \approx \qquad + 2\Delta x [2 \, \rho_{\mathsf j, \mathsf N } - 1] \frac{1}{a} U_{\mathsf j, \mathsf  N} + U_{\mathsf j, \mathsf N-1}.
\end{align}

The off-grid values $U_{\mathsf j,-1}$ and $U_{\mathsf j,\mathsf N+1}$
correspond to the ghost grid points and are eliminated by substitution from the recursion relations~\eqref{eq:forward_euler} at $\mathsf i = 0$ and $\mathsf i=\mathsf N$,
which we write for convenience:
\begin{align}
    U_{\mathsf j+1, 0} &= U_{\mathsf j, 0} + P_0 \,\left[  a^2 \frac{U_{\mathsf j, -1} - 2 U_{\mathsf j, 0} + U_{\mathsf j, 1}}{\Delta x^2} - U_{\mathsf j, 0} \right] \Delta t, \label{eq:euler_left}\\
    U_{\mathsf j+1, \mathsf N} &= U_{\mathsf j, \mathsf N} + P_{\mathsf  N} \, \left[ a^2 \frac{U_{\mathsf j,\mathsf  N-1} - 2 U_{\mathsf j,\mathsf  N} + U_{\mathsf j, \mathsf N+1}}{\Delta x^2} - U_{\mathsf j, \mathsf N}\right] \Delta t.
\end{align}
The approximated second derivatives in the square brackets therefore are
\begin{equation}
    \frac{U_{\mathsf j, -1} - 2 U_{\mathsf j, 0} + U_{\mathsf j, 1}}{\Delta x^2} = \frac{L_{\mathsf j} \, U_{\mathsf j,0} +2 U_{\mathsf j,1} }{\Delta x^2}, \quad \mbox{with}  \,
     L_{\mathsf j} = -2 - 2[2 \, \rho_{\mathsf j, 0} - 1] \Delta x \frac{1}{a},
     \label{eq:left_boundary}
\end{equation}
and
\begin{equation}
   \frac{U_{\mathsf j, \mathsf N-1} - 2 U_{\mathsf j,\mathsf  N} + U_{\mathsf j, \mathsf N+1}}{\Delta x^2} = \frac{ 2 U_{\mathsf j, \mathsf  N-1} + R_{\mathsf j} \, U_{\mathsf j, \mathsf  N}}{\Delta x^2}, \quad \mbox{with}\,
       R_{\mathsf j} = -2 + 2[2 \, \rho_{\mathsf j, \mathsf N}  -1] \Delta x \frac{1}{a},
\end{equation}
for the left and the right boundaries,
respectively.

The recruitment of new PolII at the 5' end of the gene is arrested upon chemical perturbation of the promoters with Trp.
This is modelled by assuming that a PolII  molecule at the leftmost site can leave
its position, whilst no molecules can be injected. 
Therefore, in mean-field approximation, the average number of particles at the left boundary of the system  obeys
\begin{equation}
\frac{d \phi_0(t) }{dt} = - p_0 \phi_0(t) (1 -  \phi_1(t)) ,
\end{equation}
whose hydrodynamic limit is approximated by
\begin{align}
\frac{\partial }{\partial t} \varrho(0,t) \approx& - a \, \tilde{p}(0) \varrho(0,t) (1-  \varrho(0,t) -  a \frac{\partial}{\partial x} \varrho(x,t)|^{+}_{x=0}  ).\\
\end{align}
with discretised iterative solution
\begin{equation}
  \begin{aligned}
    \rho_{j+1,0} - \rho_{j,0} & = - \Delta t \, a\, P_0 \, \rho_{j,0} \left(1-\rho_{j, 0} - a \frac{\rho_{j,1} - \rho_{j,0}}{\Delta x}  \right),\\
    \rho_{0,0} &=  \varrho_{L}(0),\\
    \rho_{0,1} &= \varrho(\Delta x,0).
    \label{eq:no_influx}
  \end{aligned}
\end{equation}
Using equations~\eqref{eq:no_influx} to iteratively update  $L_{\mathsf j}$ in \eqref{eq:left_boundary} yields
open boundary conditions, which are used in the simulation experiments. The dynamics of the hydrodynamic limit approximate well 
the occupation density a TASEP with open boundaries on a large yet finite lattice ($N=600$) as illustrated in Fig.~\ref{fig:hydro_vs_TASEP}.

\begin{figure}
    \centering
    \includegraphics[width=1\linewidth]{{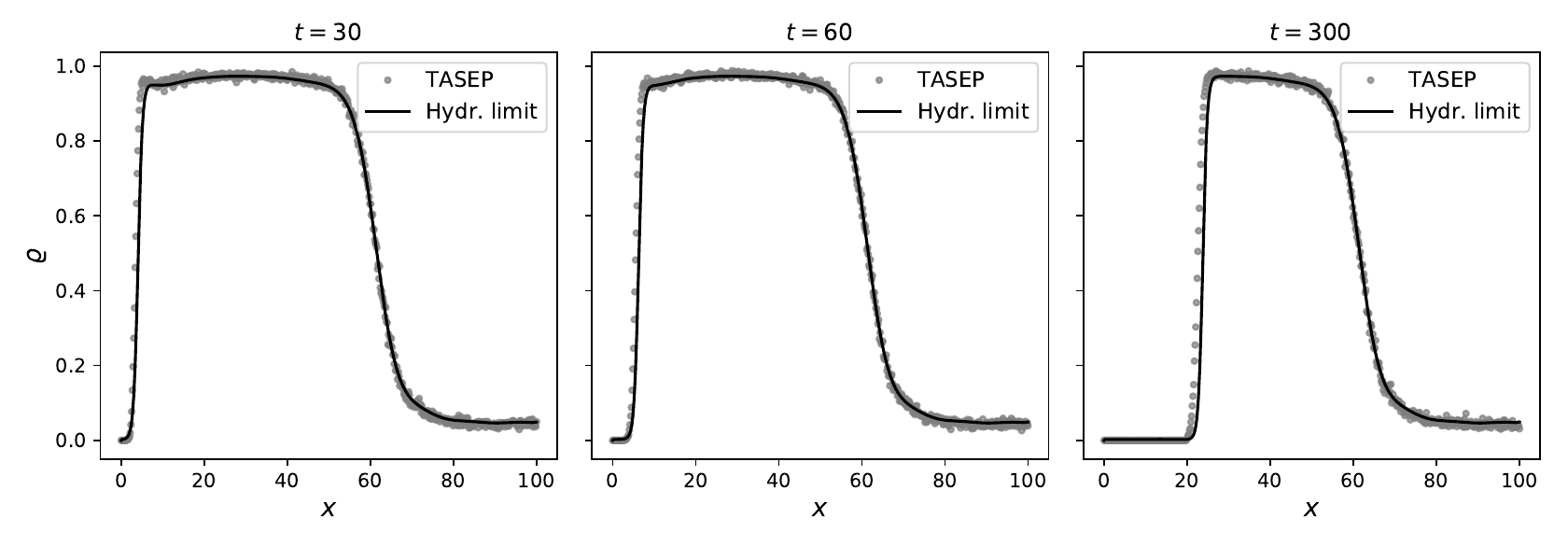}}
    \caption{
    Comparison of the hydrodynamic solution (solid lines) and Monte Carlo simulation (Gillespie algorithm as update rule, ensemble size of $1000$) occupation density of a TASEP with open boundaries, lattice size $N=600$, and smoothly varying hopping rates (grey markers). Parameters and boundary conditions as in Fig.~\ref{fig:NESS}-D.
    }
    \label{fig:hydro_vs_TASEP}
\end{figure}

Sequencing data shows that, at the beginning of the flanked region upstream of the TSS, 
 expression is very low and thus signal will remain constantly close to zero during the time course.
The reads downstream the TES are also assumed to be constant within measurement errors, given that
the Trp perturbation at the TSS does not propagate up until the TES in the longest time course.
Based on this, we set fixed left and right boundary conditions.
The grid constants $\Delta t$ and $\Delta x$ are chosen such that $\Delta t / \Delta x^2 = 1 / \tilde{p}_{\mathrm{max}}$,
which guarantees numerical stability at each grid point provided that $\tilde{p}(x) < \tilde{p}_{\mathrm{max}}$ $\forall x \in [0,L]$.
The initial values over this grid at $\mathsf{j}=0$ are obtained from the binned read profile $y^*$
 by means of linear interpolation.

\begin{figure}
\includegraphics[width=0.40\linewidth]{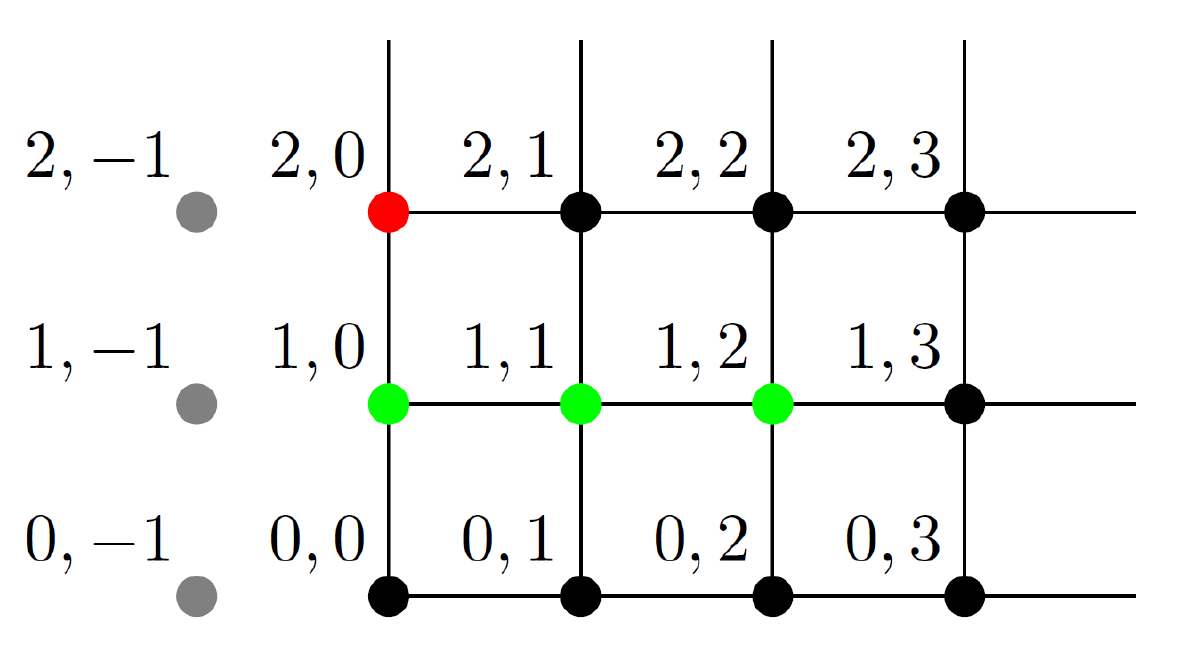}
\caption{\label{fig:stencil}
Evaluation of the discretised field $U$ at left-boundary grid nodes.
The value at the red node is obtained from the values at the green nodes. Grey points are ghost grid nodes.
}
\end{figure}

\section{MCMC sampling}
\label{app:MCMC}
\newcommand\ttilde[1]{{#1}'}

In equation~\eqref{eq:posterior} of the main text, the posterior 
probability $P(\mathbf{f},m,\sigma_\epsilon,\kappa,\sigma_f,l|\mathbf{y})$ is expressed as proportional
to the likelihood  $P(\mathbf{y}|\mathbf{f},\sigma_\epsilon,\kappa)$, multiplied by the latent GP prior $P(\mathbf{f}|m,\sigma_f,l)$ times the prior $P(m,\sigma_\epsilon,\kappa,\sigma_f,l)$. The meanings of each parameter are summarised in Table \ref{tab:parameters}.
\begin{table}[]
    \begin{tabular}{c|c|l}
    \hline
    Parameter & Prior & Description \\
     \hline
     % $\tilde{p}$ & $\tilde{p}_{\mathrm{max}}/(1+\exp(-f)), \, f\sim$  &  \\
    $\tilde{p}$ & \begin{tabular}[c]{@{}l@{}} $\tilde{p}_{\mathrm{max}}/(1+\exp(-\mathbf{f})), $ \\ $\mathbf{f}\sim\mathcal{N}(m,\mathbf{K}(\sigma_f,l)) $ \end{tabular} & \begin{tabular}[c]{@{}l@{}}Rate profile,  main target for inference and linked\\
    to a Gaussian process (GP) prior; evaluated at specific\\ points, the GP is equivalent to a multivariate normal.\end{tabular} \\[6ex]
    $m$  & $m_{\mathrm{min}}  + \frac{(m_{\mathrm{max}} - m_{\mathrm{min}})}{1+\exp(-\xi_m)}, \, \xi_m\sim\mathcal{N}(0, 1)$  &  Location parameter (mean) of the GP prior.  \\[5ex]
    $\sigma_f$  & ${\sigma_f}_{\mathrm{min}}  + \frac{({\sigma_f}_{\mathrm{max}} - {\sigma_f}_{\mathrm{min}})}{1+\exp(-\xi_f)}, \, \xi_f\sim\mathcal{N}(0, 1)$ & Amplitude of the GP prior.   \\[5ex]
    $l$  &  $l_{\mathrm{min}}  + \frac{(l_{\mathrm{max}} - l_{\mathrm{min}})}{1+\exp(-\xi_l)}, \, \xi_l\sim\mathcal{N}(0, 1)$   & \begin{tabular}[c]{@{}l@{}}Characteristic length scale of the GP prior\\ (as $l$ decreases, the GP varies more rapidly with $x$).\end{tabular}\\[5ex]
    $\sigma_\epsilon$ & ${\sigma_\epsilon}_{\mathrm{min}}  + \frac{({\sigma_\epsilon}_{\mathrm{max}} - {\sigma_\epsilon}_{\mathrm{min}})}{1+\exp(-\xi_\epsilon)}, \, \xi_\epsilon \sim\mathcal{N}(0, 1)$ & Standard deviation of normal measurement error. \\[5ex]
    $\kappa$ &  $\kappa_{\mathrm{min}}  + \frac{(\kappa_{\mathrm{max}} - \kappa_{\mathrm{min}})}{1+\exp(-\xi_\kappa)}, \, \xi_\kappa\sim\mathcal{N}(0, 1)  $ & 
    \begin{tabular}[c]{@{}l@{}} Proportionality factor to convert ChIP-seq reads\\ into occupation density. \end{tabular} \\[5ex]
    \hline
    \end{tabular}
    \caption{ Target parameters for inference.}
    \label{tab:parameters}
    \end{table}
To sample from  the  posterior we used a blocked Gibbs sampling
scheme where we recursively update the latent variable $\mathbf{f}$,
the two covariance variables $l,\sigma_f$, and the three likelihood variables  $m,\kappa,\sigma_\epsilon$, using the so-called elliptic slice sampler~\cite{Murray2010} for each block.

In order to update $\mathbf{f}$ as a first block we parametrise the posterior such that
$m$ is absorbed into the likelihood, i.e.,
\begin{equation}
     P(\ttilde{\mathbf{f}},m,\sigma_\epsilon,\kappa,\sigma_f,l|\mathbf{y}) \propto P(\mathbf{y}|\ttilde{\mathbf{f}},m,\sigma_\epsilon,\kappa)
     P(\ttilde{\mathbf{f}}|\sigma_f,l) P(m)P(\sigma_\epsilon) P(\kappa) P(\sigma_f)P(l),
\end{equation}
where $\ttilde{\mathbf{f}}=\mathbf{f}-m$, and sample from the unnormalised conditional $P(\mathbf{y}|\ttilde{\mathbf{f}},m,\sigma_\epsilon,\kappa) P(\ttilde{\mathbf{f}}|\sigma_f,l)$ with all hyperparameters $m$, $\sigma_\epsilon$, $\kappa$, $\sigma_f$, and $l$ held fixed.

Updating the second block $(l,\sigma_f)$ is difficult as this is strongly correlated with $\mathbf{f}$. Hence we adopt another parametrisation, which consists of expressing the multivariate Gaussian random variable $\ttilde{\mathbf{f}}$ as the deterministic function of $\boldsymbol{\nu}$
\begin{equation}
    \ttilde{\mathbf{f}}(\boldsymbol{\nu}, \sigma_f,l)=\mathbf{L}  \boldsymbol{\nu},
\end{equation}
where $\mathbf{L}(\sigma_f,l)$ satisfies $\mathbf{L} \mathbf{L}^{-1} = \mathbf{K}(\sigma_f,l)$ (it can be computed from $\mathbf{K}$ using a Cholesky decomposition) and $\boldsymbol{\nu} \sim \mathcal{N}(0,\mathbbm{1})$ is independent of the other variables. 
This parametrisation incorporates all variables into the likelihood and results in the equivalent posterior
\begin{equation}
     P(\boldsymbol{\nu},m,\sigma_\epsilon,\kappa,\sigma_f,l|\mathbf{y}) \propto P(\mathbf{y}|\ttilde{\mathbf{f}}(\boldsymbol{\nu},\sigma_f,l),m,\sigma_\epsilon,\kappa)
     P(\boldsymbol{\nu})P(m)P(\sigma_\epsilon) P(\kappa) P(\sigma_f) P(l).
     \label{eq:equiv_posterior}
\end{equation}
For a given value of $(\ttilde{\mathbf{f}}, l,\sigma_f)$, $\nu$ is obtained as
$\nu={\mathbf{L}(\sigma_f,l)}^{-1} \, \ttilde{\mathbf{f}}$. Hence with parameters $\ttilde{\mathbf{f}},m,\kappa$, and $\sigma_\epsilon$ held fixed, we update $(\sigma_f, l)$ with a sample from the unnormalised conditional $P(\mathbf{y}|\ttilde{\mathbf{f}}(\boldsymbol{\nu},\sigma_f,l),\kappa,\sigma_\epsilon)P(\sigma_f)P(l)$.

In the third step, given the values for $\ttilde{\mathbf{f}}$, $l$, and $\sigma_f$, the likelihood hyperparameters $m$, $\sigma_\epsilon$, and $\kappa$ are updated by sampling from the unnormalised conditional $P(\mathbf{y}|\ttilde{\mathbf{f}}, m, \sigma_\epsilon)P(m)P(\sigma_\epsilon)P(\kappa)$.
We refer the reader to~\cite{Tegner2019} for more details.
  Posterior predictions are in very good agreement with time-course observations. Concentration of posterior-predictive samples around target data points
can be appreciated by zooming in (Fig.~\ref{fig:posterior_zoomin}). 
\begin{figure}
    \centering
    \includegraphics[width=0.80\linewidth]{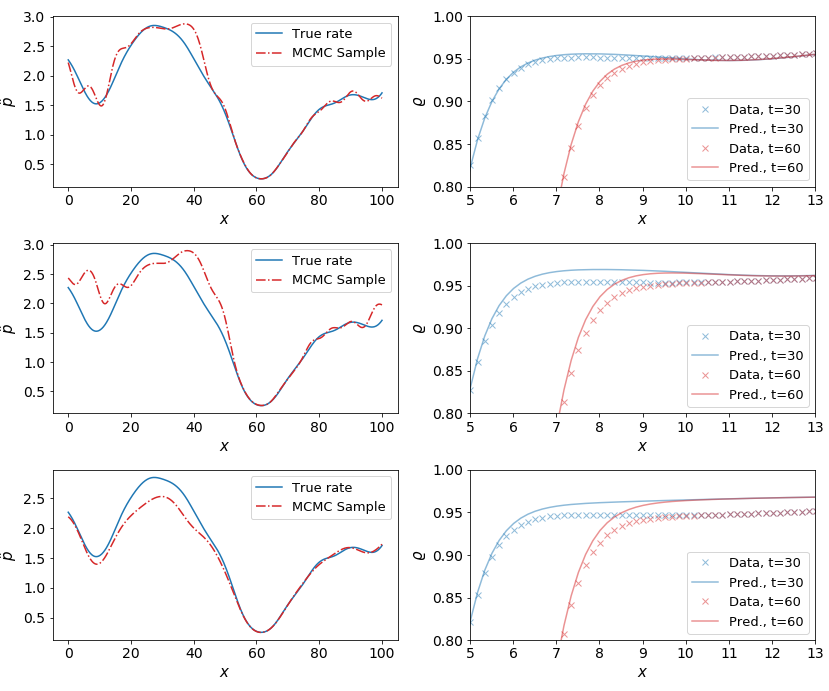}
    \caption{
        \label{fig:posterior_zoomin}
        The rate profile samples sometimes have jagged profiles (left).
        The resulting density profiles are  smoother  than the corresponding rates and their dispersion around the input data profile can be appreciated by zooming in the plot of Fig.~\ref{fig:NESS}-D (right).
    }
\end{figure}
The model is weakly sensitive to changes in the rate profile except for the bottleneck minimum at $x\approx 60$, which is sampled with high confidence.

\section{Supplemental figures}

Figs.~\ref{fig:s1_pos_predict_}, \ref{fig:s2_pos_predict_}, and  \ref{fig:s3_pos_predict_} illustrate the posterior predictive samples and the scaled ChIP-seq reads.

\begin{figure}
    \includegraphics[width=0.9\linewidth]{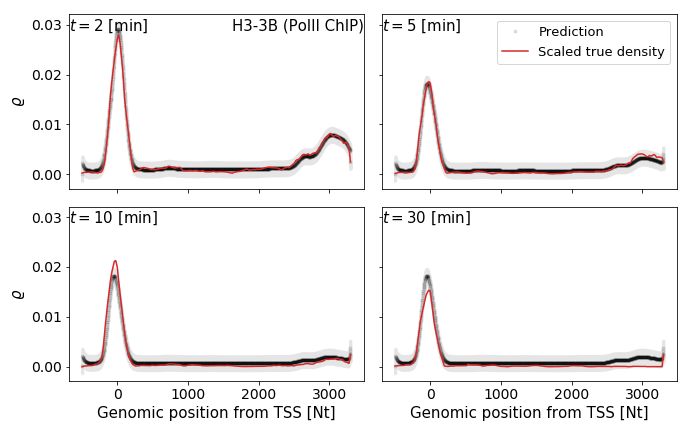}
    \caption{
    \label{fig:s1_pos_predict_}
        Predicted density profiles for gene H3-3B from PolII ChIP-seq, $2$, $5$, $10$, and $20$ minutes after Trp treatment. Markers are posterior predictive samples. Grey area is 95\% credible interval due to noise model.}
\end{figure}

\begin{figure}
    \includegraphics[width=0.9\linewidth]{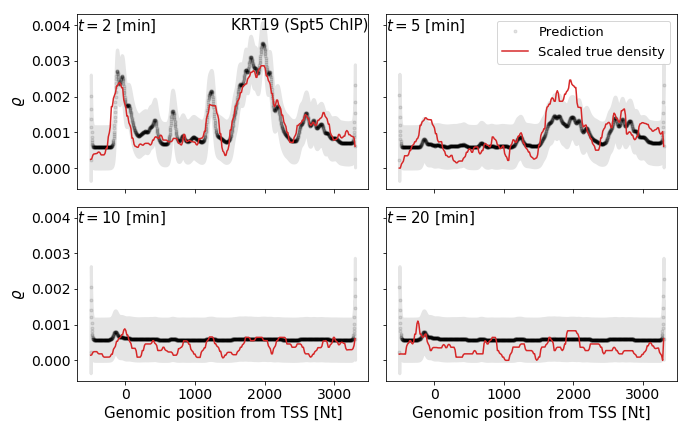}
    \caption{
    \label{fig:s2_pos_predict_}
        Predicted density profiles for gene KRT19 from Spt5 ChIP-seq. Figure keys as in Fig.~\ref{fig:s1_pos_predict_}.}
\end{figure}

\begin{figure}
    \includegraphics[width=0.9\linewidth]{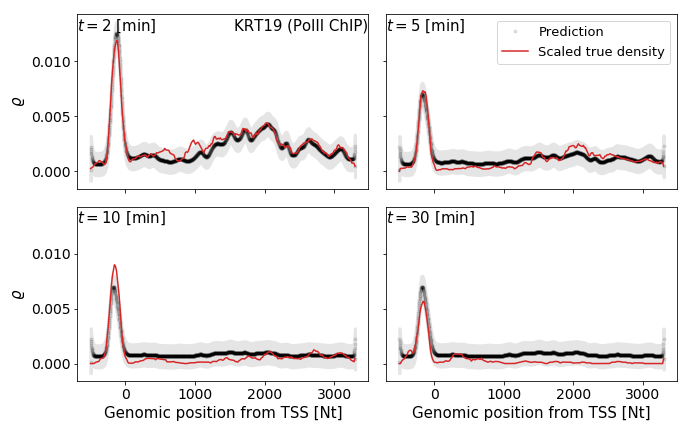}
    \caption{
    \label{fig:s3_pos_predict_}
           Predicted density profiles for gene KRT19 from PolII ChIP-seq. Figure keys as in Fig.~\ref{fig:s1_pos_predict_}.}
\end{figure}

\end{widetext}
\end{document}